\documentclass[12pt,preprint]{aastex}

\begin{document}

\title{$VI$ PHOTOMETRY OF GLOBULAR CLUSTERS NGC~6293 AND NGC~6541:
THE FORMATION OF THE METAL-POOR INNER HALO GLOBULAR CLUSTERS\footnotemark[1]}

\author{Jae-Woo Lee\altaffilmark{2} \& Bruce W.\ Carney\altaffilmark{3}}

\footnotetext[1]{Based on Observations made with the NASA/ESA
{\it Hubble Space Telescope}, obtained at the Space Telescope
Science Institute, which is operated by AURA, Inc.,
under NASA contract NAS 5-26555.}

\altaffiltext{2}{Department of Astronomy and Space Science,
Astrophysical Research Center for the Structure and
Evolution of the Cosmos,
Sejong University, 98 Gunja-Dong, Gwangjin-Gu, Seoul, 143-747, Korea;
jaewoolee@sejong.ac.kr}
\altaffiltext{3}{Department of Physics \& Astronomy,
University of North Carolina, Chapel Hill, NC 27599-3255;
bruce@unc.edu}

\begin{abstract}
We present $VI$ photometry of the metal-poor inner halo globular clusters
NGC~6293 and NGC~6541 using the planetary camera of the WFPC2 on board
{\em Hubble Space Telescope} (HST).  Our color-magnitude diagrams
of the clusters show well-defined blue horizontal branch (BHB)
populations, consistent with
their low metallicities and old ages. NGC~6293 appears to have blue straggler
stars in the cluster's central region.
We discuss the interstellar reddening and the distance modulus
of NGC~6293 and NGC~6541 and obtain $E(B-V)$ = 0.40 and $(m-M)_0$ = 14.61
for NGC~6293 and $E(B-V)$ = 0.14 and $(m-M)_0$ = 14.19 for NGC~6541.
Our results confirm that NGC~6293 and NGC~6541 are clearly located
in the Galaxy's central regions ($R_{GC} \leq$ 3 kpc).
We also discuss the differential reddening across NGC~6293.
The interstellar reddening value of NGC~6293 appears to vary by
$\Delta E(B-V)$ $\approx$ 0.02 -- 0.04 mag within our small field of view.

The most notable result of our study is that the inner halo clusters
NGC~6293 and NGC~6541 essentially have the same ages as M92,
confirming the previous result
from the HST NIC3 observations of NGC~6287.
\end{abstract}


\section{INTRODUCTION}
Understanding the formation of our Galaxy has always been one of the key quests
in modern astrophysics for decades. During the last decade,
a tremendous amount of observational data have accumulated regarding
the formation of our Galaxy, in particular due to the advent of HST.
However, questions associated with the formation and early evolution of
our Galaxy remain unanswered.

One of the earliest models for the formation of the Galaxy was
that of Eggen, Lynden-Bell, \& Sandage (1962), who postulated a
rapid, monolithic ``collapse" of a proto-Galaxy. Using similar data,
Isobe (1974) and Saio \& Yoshii (1979) suggested the formation
process could have been much longer than a free-fall timescale,
occurring over several billion years. An alternate view of the
halo formation process was presented by Searle \& Zinn (1979).
Under the assumption that variations in the numbers of blue and
red horizontal branch stars in globular clusters is an age
indicator, they argued for somewhat more youthful clusters at
larger Galactocentric distances. If true, they argued that the
Galactic halo may have formed from accretion of ``proto-Galactic
fragments", implying a more chaotic view of the halo's formation.
Carrying the argument further,
Zinn (1993) argued that the bulk of the old halo globular clusters
formed during a monolithic dissipative collapse while
the young halo globular clusters formed independently of the Galaxy
and later accreted into our Galaxy.
The longer timescale of a monolithic formation process and
the fragmentary accretion model both suggest a variation in
the ages of globular clusters, and a natural assumption is
that the Galactic halo may have formed ``inside-out", with star
formation beginning earlier in center due to
the smaller free collapse time scale ($\tau \propto \rho^{-1/2}$).

The discoveries that the Universe's expansion and self-gravity are
dominated by dark energy and dark matter have led to the
basic ``$\Lambda$CDM" model, and the formation of galaxies
within this model is a much more elaborate yet tractable
variant of the original Searle \& Zinn (1978) hypothesis,
and that of White \& Rees (1978). Numerous sub-galactic mass
concentrations form rapidly in the early Universe, and undergo
mergers to form the larger galaxies we see today (see
Navarro, Frenk, \& White 1997). Mergers
are still going on in the current epoch, in our own Milky Way
Galaxy, and in many others as well. Qualitatively, this
``hierarchical assembly" model implies an ``inside-out"
formation of our Galaxy, and detailed calculations show this
to be the case for the dark matter (Helmi et al.\ 2003)
and the stars as well (Robertson et al.\ 2005; Font et al.\ 2006).

What do we know about how when star formation began and how
rapidly it proceeded in these fragments and the young Galaxy?
Two techniques are available to answer these questions.
Element-to-iron (or other elemental) ratios provide valuable
clues due to different elements' different nucleosynthesis
sites. For example, light elements such as magnesium, calcium,
and silicon are manufactured more readily in short-lived stars
that explode as Type II supernovae, whereas iron is more readily
synthesized in Type Ia supernovae, which are thought to arise from
mass transfer or mergers of white dwarfs, a process which is
thought to require a longer period of time, perhaps $10^{9}$
years or more. Thus in a closed system of stars and gas,
stars with high [$\alpha$/Fe] abundance
ratios were likely born prior to significant contributions
from Type Ia supernovae. The discovery that the Galaxy's
dwarf spheroidal galaxy neighbors have unusually low
[$\alpha$/Fe] ratios, even at very low [Fe/H] values (Shetrone
et al.\ 2001, 2003; Venn et al.\ 2004) thereby presented
a challenge to the idea that the Galaxy's halo was assembled from
proto-Galactic fragments, since, apparently, some of the surviving
fragments have not shared the same chemical enrichment history
as the Galactic halo. The difficulty with relying on
element-to-iron ratios is that they do not monitor ages
directly. Small galaxies producing stars very slowly
will experience enrichment from Type Ia supernovae before
the overall [Fe/H] value has risen significantly. Further,
incorporation of supernovae ejecta in subsequent generations
of stars depends on the host galaxy's ability to retain gas.
Font et al.\ (2006) have shown that the $\Lambda$CDM model
is able to explain the differences in [$\alpha$/Fe] vs.\ [Fe/H]
patterns
in the local dwarf spheroidal galaxies vs.\ the Galactic halo
due to the different rates of star formation in the systems
that were absorbed by the Galaxy compared to the survivors.

The second method of age dating is the most direct: comparisons
of main sequence turn-offs with stellar model isochrones.
While the derived ages do not depend on the history of
nucleosynthesis, the method is difficult to apply with
the same level of precision. For example, if the timescale
for significant contributions of Type Ia supernovae to
manifest their presence in subsequent generations of stars,
a change in [$\alpha$/Fe] can be expected to be revealed in
stars whose ages differ by less than $10^{9}$ years
{\em in a closed system}\footnote{It is even more problematical
in ensembles constructed from merger fragments, each that
may have experienced a different star formation history.}. Measurement
of even relative ages with such precision using main sequence
data is very challenging. Nonetheless, much careful work
has already been done in this area.

We are interested in the particular question of how rapidly
did star formation begin throughout the Galaxy and its
then more numerous proto-Galactic fragment.
The HST observations of one of the most remote metal-poor halo clusters
NGC~2419 ($R_{GC}$ $\approx$ 90 kpc) of Harris et al. (1997)
have shown that NGC~2419 and M92 ($R_{GC}$ = 9.6 kpc)
essentially have the same age within $\pm$2 Gyr,
suggesting that globular-cluster formation must have started
at everywhere at about the same time in our Galaxy
(Richer et al. 1996; see also the counterargument of
Chaboyer, Demarque, \& Sarajedini 1996).
Moreover, the local dwarf galaxies Carina (Mighell 1997), Draco
(Grillmair et al. 1998), Leo I \& II (Held et al. 2000;
Mighell \& Rich 1996), and Ursa Minor (Mighell \& Burke 1999) and
the globular clusters in the Fornax dwarf galaxy
(Buonanno et al. 1998), LMC (Johnson et al. 1999), and
the Sagittarius dwarf galaxy (Montegriffo et al. 1998;
Layden \& Sarajedini 2000) appear to be coeval with
typical globular clusters of similar metallicity in our Galaxy.
These similar ages suggest that globular cluster formation
may have begun everywhere at the same time not only in our Galaxy
but also in the local dwarf galaxies,
despite the differences in the initial physical environments.

The ``inside-out" model led van den Bergh (1993)
to suggest that the most metal-poor globular cluster  near the Galactic center
NGC~6287 ([Fe/H] = $-$2.01, Lee \& Carney 2002; $R_{GC}$ = 1.6 kpc,
Lee et al.\ 2001) may be the oldest globular cluster in our Galaxy, based on
its metallicity, horizontal branch (HB) population, and its spatial location
in our Galaxy. The recent HST NIC3 observations of the cluster
(Lee et al.\ 2001) have shown that NGC~6287 and M92 essentially have the same
age within $\pm$2 Gyr, suggesting that the metal-poor globular cluster
formation must have been triggered roughly everywhere at the same time in our Galaxy.
It should be noted that, however, HST NIC3 photometry can be
vulnerable to the variation in the intrapixel sensitivity (Lauer 1999) and
the temperature dependence of the interstellar reddening law
in the HST NICMOS F110W/F160W photometric system (Lee et al.\ 2001).
Therefore, a photometric study with an expanded sample,
preferably employing detector plus filter systems not suffering such effects
as can be seen in HST NIC3, is necessary to rank the formation epoch
of metal-poor inner halo globular clusters in comparison to the intermediate
or the outer halo globular clusters.

NGC~6293 ($\alpha$ = 17:10:10.42; $\delta$ = $-26$:34:54.2;
$\ell$ = 357.62; $b$ = +7.83) and NGC~6541 ($\alpha$ = 18:08:02.20;
$\delta$ = $-43$:42:19.7; $\ell$ = 349.29; $b$ = $-11.18$)
are the second and the third most metal-poor
globular clusters within 3 kpc from the Galactic center.
NGC~6293 and NGC~6541 are located $\approx$ 1.4 kpc and 2.2 kpc from
the Galactic center and $\approx$ 1.2 kpc and $-$1.4 kpc from
the Galactic plane, suffering interstellar reddening of
$E(B-V)$ = 0.41 and 0.14, respectively (Harris 1996).
In Table~\ref{tab1}, we list clusters' properties. Since both clusters are thought to be
post core-collapsed, as can be more frequently found in the inner halo region
than other parts of our Galaxy (Barbuy, Bica, \& Ortolani 1998),
ground-based photometric study of these two clusters has been limited.

Janes \& Heasley (1991) presented $BV$ photometry of NGC~6293.
Their color-magnitude diagram (CMD) showed that the red-giant branch (RGB) and
the BHB morphologies of NGC~6293
are similar to those of M92, indicating that the cluster
is both old and metal-poor.
By comparing with M92, they suggested that $E(B-V)$ = 0.46 and
$(m-M)_0 \approx$ 16.0. Since their observations reached just to
the main-sequence turnoff (MSTO), however, they were not able to address
the age of the cluster. They also made an interesting point that
the RGB-tip magnitude of NGC~6293 appears to be a full magnitude fainter
than that of M92. They suggested that this may reflect that the cluster has
undergone core collapse. Trager et al.\ (1995) found that the cluster has
indeed undergone core collapse, as may NGC~6541. More recent work by
Noyola \& Gebhardt (2006), relying on HST WFPC2 data,
confirms that both clusters show sharply
rising surface brightness levels into the innermost regions.

Alcaino (1971, 1979) obtained $BV$ photographic photometry and
Alcaino et al.\ (1997) presented multi-color CCD photometry of NGC~6541.
Alcaino (1979) noted that NGC~6541 has BHB stars and its CMD appears to be
similar to that of M13. He also noted that NGC~6541 appears to be deficient
in bright RGB stars. With deeper CCD photometry, Alcaino et al.\ (1997)
claimed that the CMD of NGC~6541 is similar to those of M13 and M79,
finding $E(B-V)$ = 0.15 and [Fe/H] $\approx$ $-$1.8 for NGC~6541.
They noted a discrepancy in the position of NGC~6541 BHB stars
in $U - (B - V)$ and $(U - B) - (B - V)$ diagrams, in the sense that
$U$ magnitude of BHB stars in NGC~6541 appear to be $\approx$ 0.3 -- 0.4 mag
brighter than those in M79. Using an isochrone fitting method, they claimed
that NGC~6541 is very old, one of the oldest clusters they had studied.

Lee \& Carney (2002) presented high-resolution echelle
spectroscopy of both clusters. They noted that both clusters appear
to be silicon enhanced and titanium depleted compared to the
intermediate halo clusters. In particular, they suggested that
[Si/Ti] ratios appear to be related to Galactocentric distances,
in the sense that [Si/Ti] ratios decrease with Galactocentric distance,
and they proposed that these [Si/Ti] gradients with Galactocentric distance
may be due to the different masses of the SNe II progenitors.
The high [Si/Ti] ratios toward the Galactic center may be due to
higher-mass SNe II contributions. This is an additional hint that
both clusters may have formed particularly early.
Lee \& Carney (2002) also derived the radial velocity of the clusters
(see also Table~\ref{tab1}). Their results suggest that the radial velocities of
NGC~6293 and NGC~6541 are small enough that they are most likely
genuine inner halo clusters unless their tangential motions are
extremely large. Unfortunately, neither cluster has a measured
proper motion as yet.

In this paper, we explore the relative ages of inner halo clusters
NGC~6293 and NGC~6541 using HST PC2 photometry.
As mentioned earlier, both clusters have high central concentrations
so that only HST can provide the necessary high angular resolution
to study main-sequence photometry of these clusters.
In section 2, data acquisition and data reduction are discussed.
We present the CMDs of the clusters and discuss the main sequence fittings
of the clusters to M92 and NGC~2419 in section 3. The interstellar reddening
and the distance modulus are also discussed in this section.
Finally, we discuss the formation epoch of the metal-poor inner halo
clusters in section 4.

\section{OBSERVATIONS AND DATA REDUCTION}

The observations of NGC~6293 and NGC~6541 were carried out on 23 March 1995
(UT) and 15 October 1994 (UT), respectively. HST PC2 is equipped with
a 800 $\times$ 800 pixel Loral CCD. The FOV is 36 $\times$ 36 arcsec,
and which gives 0.0455 arcsec/pixel (f/28.3). For both clusters images at
two different locations with a small overlap region were obtained.
For NGC~6293, the locations of the pointings were
$\alpha$ = 17:10:12.55,
$\delta$ = $-26$:33:58.66 (J2000; NGC~6293-F1) and
$\alpha$ = 17:10:11.06,
$\delta$ = $-26$:35:18.66 (NGC~6293-F2).
The integration times were 40 sec and 500 sec for the F555W passband and
60 sec and 700 sec for the F814W passband.
For NGC~6541, the locations of the pointings were
$\alpha$ = 18:08:06.30,
$\delta$ = $-43$:41:51.21 (NGC~6541-F1) and
$\alpha$ = 18:08:04.46,
$\delta$ = $-43$:42:11.21 (NGC~6541-F2).
The integration times were 12 sec and 140 sec for the F555W passband and
20 sec and 260 sec for the F814W passband.
No problems were reported during the observations.
Table~\ref{tab2} lists the journal of observations.

The raw images were calibrated using the HST standard calibration pipeline
to perform bias subtraction, dark current subtraction, flat field correction,
and shutter shade correction (WFPC2 Group 1998).
The final processed F555W images for NGC~6293 and NGC~6541 are presented
in Figure~1 and Figure~2, respectively. With the calibrated images,
we masked the vignetted portions of the PC chip, which are not usable
for further analysis.

To perform point-spread function (PSF) photometry for all calibrated images,
we used  DAOPHOTII, ALLSTAR and ALLFRAME (Stetson 1994, 1995; Turner 1995)
following the procedure described in Lee \& Carney (1999a)
and Lee et al.\ (2001). With the PSF magnitudes returned from the final
ALLFRAME run, we performed the aperture corrections. Holtzman et al.\ (1995a)
adopted aperture magnitudes with a 0.5 arcsec ($\approx$ 11 pixel in HST PC)
aperture size and the sky annuli of 4 arcsec ($\approx$ 88 pixel) to 6 arcsec
($\approx$ 132 pixel) in deriving their transformation relations between
the HST F555W/F814W photometric system and the standard Johnson-Cousins $VI$
photometric system. Johnson et al.\ (1999) claimed that, however, the sky
annuli that Holtzman et al.\ (1995a) used are too large for crowded field
photometry and they adopted sky annuli of 2.0 arcsec ($\approx$ 44 pixel) and
2.5 arcsec ($\approx$ 55 pixel). They suggested that this reduces the effects
of badly subtracted neighbors in crowded regions and uneven sky brightness
across the clusters. They noted that the magnitude difference between using
the larger and the smaller sky annuli is only 0.001 mag and it appears to
be much smaller than the errors induced by other effects.
We compare the aperture magnitude differences using the two different sizes
of the sky annuli in NGC~6541-F1, which is the least crowded
among our fields, and we obtain 0.002 $\pm$ 0.015 mag.
The difference appears to be negligible and therefore we adopt a 0.5 arcsec
aperture size and the sky annuli of 2.0 arcsec and 2.5 arcsec during
our aperture corrections.

The HST WFPC2 cameras introduce a geometric
distortion as a result of their optical design (Holtzman et al.\ 1995b).
The geometric distortion causes the effective pixel area to vary across
the CCD chip and affects astrometry and the point source photometry.
The pixel area effect is corrected during the flat field correction and
consequently, after flattening, all pixels are normalized to the same area.
Photometrically, this preserves surface brightness but causes a change
in the pixel scale with the biggest effect at the edges of the chip
(see, for example, Figure~15 of Holtzman et al.\ 1995b).
Thus, the geometric distortion does not affect surface photometry,
but does alter point source photometry.
We corrected the geometric distortion effect in our images using
the prescription of WFPC2 group (1998).

The HST WFPC2 CCDs also have a significant charge transfer efficiency (CTE)
problem which causes some signal to be lost when charge is transferred
down the chip during readout. This has the effect of making objects at
higher row numbers appear fainter than they would if they were at low row
numbers because more of the charge gets trapped for the objects near the top
of the chip. The effect also depends on the temperature of the chip,
the brightness of the sources, in the sense that the CTE effect is larger
for the faint sources, and the amount of background charge on the chip,
such that there is significantly less CTE effect due to the enhanced pedestal
charge level of the chip when the background is bright.
Hence the CTE effect on the PC2 chip is expected to be larger than
those on the WF2 chips due to the lower background.
We corrected the CTE problem in our images using the prescription
of Whitmore \& Heyer (1997).

Finally, the transformation to the standard Johnson-Cousins $VI$ photometric
system was performed using the relations given by Holtzman et al.\ (1995a).
Following Holtaman et al.\ (1995a), the calculations were performed
iteratively and $V$ magnitudes and $(V-I)$ colors converged to
$\pm$ 0.001 mag within 3 iterations.

In Figure~3, we show the differences in $V$ magnitudes
and $(V-I)$ colors as functions of $V$ magnitudes of the stars
in the overlap regions in NGC~6293 and NGC~6541. The residuals are
in the sense of the field F1 minus the field F2 in each cluster.
For our calculations, we used stars with $V$ $\leq$ 23.0 mag
for NGC~6293 and $V$ $\leq$ 22.5 mag for NGC~6541 and we excluded the stars
detected within 10 pixels from the edges of the chip. As shown in the Figure,
the magnitudes and colors are in good agreement between the two fields
in each cluster with $\Delta V$ = 0.003 $\pm$ 0.006 mag
and $\Delta (V-I)$ = 0.001 $\pm$ 0.003 mag for NGC~6293 (431 stars) and
$\Delta V$ = $-$0.002 $\pm$ 0.005 mag and
$\Delta (V-I)$ = 0.007 $\pm$ 0.004 mag for NGC~6541 (122 stars),
suggesting that the zero-point adjustments are not necessary
in combining photometry of the two fields. (The errors are those of the mean.)
Therefore, zero-point offsets were not applied in our results.

\section{RESULTS AND DISCUSSIONS}

\subsection{Color-Magnitude Diagrams}

In Figure~4, we present CMDs of NGC~6293 for the stars in the fields F1
(6761 stars), F2 (3150 stars), and the overlap region (844 stars).
Our CMDs for NGC~6293 show a well-developed BHB morphology, in particular
in the central part of the cluster (the field F1), confirming the previous
result of the ground-based $BV$ photometry of Janes \& Heasley (1991).
As discussed by Lee \& Carney (2002), NGC~6293 has
a low metallicity ([Fe/H] = $-$1.99) and its BHB morphology is consistent with
its metal-poor nature. Since the RGB stars brighter than $V$ $\approx$ 14.5 mag
were saturated even in the short exposure frames during the observations,
we are not able to test the
Janes \& Heasley (1991) observation that the RGB tip is
a full magnitude fainter than that of M92.

Figure~5 shows CMDs of NGC~6541 for the stars in the fields F1 (1086 stars),
F2 (2162 stars), and the overlap region (235 stars).
Our CMDs show the BHB population in NGC~6541, confirming the previous
result of Alcaino (1979). Our CMDs may support the idea (Alcaino 1979)
that NGC~6541 is deficient in luminous RGB stars.
Since the FOV of the HST PC is small and the two fields are located
78 arcsec and 53 arcsec from the cluster center, however,
the lack of RGB stars in NGC~6541 may not be conclusive.

Our CMD data for NGC~6293 and NGC~6541 are available upon request to
the authors or the electronic version of the journal.
Sample CMD data for NGC~6293 are shown in Table~\ref{tab3}.
Note that the first column of Table represents the field identification
(``1" for the field F1 only, ``2" for the field F2 only, and
``o" for the overlap region) and the positions (columns 3 and 4) are presented
in the pixel coordinate unit of the detector (0.0455 arcsec pixel$^{-1}$).

\subsection{Blue Stragglers}

Blue straggler stars
(BSS) are located in the region between the MSTO and the BHB.
We found 22 BSS candidates in NGC~6293 (19 BSS candidates in the field F1 only,
1 in the field F2 only, and 2 in the overlap region).
Since the observations did not cover off-cluster fields,
we are not able to correct for field star contamination in our CMDs.
However, the BSS candidates in NGC~6293 appear to be more centrally
concentrated (i.e.\ the frequency of the BSS candidates is higher
in the field F1 than in the field F2) and, therefore, they seem more likely
related to the cluster rather than the off-cluster field populations.
Are the numbers of blue stragglers enhanced by the dynamical state of
a cluster? While this is plausible, in that binary star orbits should
harden by interactions in dense environments, evidence to support
the conjecture is hard to find. Ferraro et al.\ (1995), Davies et al.\ (2004),
Piotto et al.\ (2004), and Sandquist (2005) do not support the idea.
We have followed the approach of Piotto et al.\ (2004), and compared
the numbers of blue stragglers (22) to horizontal branch stars (74).
Therefore, $F_{\rm BSS}$ = 0.297, and log~$F_{\rm BSS}$ = $-0.527$.
NGC~6293 does not seem out of place in Figure~1 of Piotto et al.\ (2004),
where log~$F_{\rm BSS}$ is compared to $M_{\rm V}$ ($-7.77$) and
the logarithm of the central density (log~$\rho$ = 5.22).

\subsection{Main-Sequence Fitting And The Relative Age Estimates}

We obtain the mean loci of the clusters by using a combination of subjective
removal of outliers followed by use of iterative sigma-clipping mean color
calculations. The mean loci of the clusters are presented in Tables~\ref{tab4} and \ref{tab5}
for NGC~6293 and NGC~6541, respectively. Note that the fifth and
the tenth columns of the Tables represent the number of stars used
in our calculations, not the clusters' luminosity functions.
The $V$ magnitudes and $(V-I)$ colors of the MSTO are 20.01 mag and 1.05 mag
for NGC~6293 and 18.81 mag and 0.75 mag for NGC~6541.
Our MSTO magnitude and color of NGC~6541 agree well with those
of Alcaino et al.\ (1997), who obtained $V$ = 18.9 $\pm$ 0.1 mag and
$(V-I)$ = 0.750 mag for the NGC~6541 MSTO.

We explore the relative ages of NGC~6293 and NGC~6541 with respect to M92
using, in essence, color differences between the base of the RGB (BRGB) and
the MSTO. VandenBerg, Bolte, \& Stetson (1990) described a method to derive
accurate relative cluster ages with similar chemical compositions.
They recommended shifting the clusters' CMDs in color until the TO colors
agreed, then shifting the magnitudes until the upper main-sequence is matched
at a point 0.05 mag redder than the TO. The color differences of the resulting
BRGB could then be used to estimate the relative age of the clusters,
since the color difference between the TO and the BRGB is a monotonic and
inverse function of age. Further, this color difference is independent of
differences in cluster distances and reddenings,
as well as differences in photometric zero-points.
This method is known to be nearly insensitive to metallicity within moderately
large ranges in metallicity (see below).
However, it should be emphasized that this method is vulnerable to
the oxygen abundance. Since oxygen has a higher ionization potential
than the other abundant heavy elements (magnesium, silicon, calcium, iron),
the difference in the oxygen abundance affects the color and the luminosity
of TO stars but not those of RGB stars, in the sense that higher oxygen
abundance makes a redder and less luminous TO. Oxygen also plays a vital role
in the hotter part of CNO bi-cycle, so the TO luminosity is also affected
(VandenBerg 1992). VandenBerg \& Stetson (1991) studied the effect of the oxygen
abundance on the relative age estimations using this method
(see also VandenBerg \& Bell 2001). Their results suggested that changes
in the oxygen abundance could mimic age differences, in the sense that
the higher oxygen abundance imitates an older age at a given metallicity.
However, VandenBerg \& Stetson (1991) also pointed out that this effect is
less important at the lowest metallicities and at the oldest ages.

Lee \& Carney (2002) discussed the elemental abundance analyses
of two RGB stars in NGC~6293 and NGC~6541.
The metallicities of NGC~6293 and NGC~6541 are [Fe/H] = $-$1.99 and
$-$1.76 from the ionized iron lines and they appear to be
0.2 and 0.45 more metal-rich than M92.
The mean $\alpha$-element abundances for NGC~6293 and NGC~6541 are
[$\alpha$/Fe] = +0.32 and +0.36, slightly lower than that of M92,
[$\alpha$/Fe] = +0.41. Using the relation given by Chieffi et al.\ (1991),
we obtained the overall metallicity $Z$ = 0.0003 for NGC~6293 and 0.0006
for NGC~6541, respectively, and they are larger than that of M92 ($Z$ =0.0002).
Therefore, a slight metallicity effect in the relative age estimate would be
expected, in particular for NGC~6541 (see below).
The original stellar oxygen abundances are hard to determine
for highly-evolved RGB stars since
oxygen appears to be depleted due to internal mixing.
We obtained [O/Fe] = $-$0.14 from one RGB star in NGC~6293 and
$-$0.58 and $-$0.09 from two RGB stars in NGC~6541.
(We are not able to determine the oxygen abundance from one RGB star
in NGC~6293, because its oxygen lines are too weak to be measured reliably.)
A low sodium abundance ([Na/Fe] $\leq$ 0) appears to be a good indicator
of minimal mixing in globular cluster RGB stars
(Langer, Hoffman, \& Sneden 1993; Kraft 1994).
The sodium abundances of the RGB stars in NGC~6293 and NGC~6541 are
[Na/Fe] $\gtrsim$ +0.3, suggesting that they have experienced some mixing.
The oxygen abundances of the clusters should be determined using
an expanded sample at fainter magnitudes in the future.
For the present, we assume that the initial [O/Fe] ratios for NGC~6293,
NGC~6541, and M92 are very similar ([O/Fe] $\approx$ [$\alpha$/Fe]) and,
therefore, the TO luminosities of the clusters are not affected by
the difference in the oxygen abundance.

Figure~6 plots $(V-I) - (V-I)_{TO}$ versus $V - V_{+0.05}$
of enhanced $\alpha$-element ([$\alpha$/Fe] = +0.3) model isochrones
(Bergbusch \& VandenBerg 2001) with different metallicities
at the age of 14 Gyr to illustrate the metallicity dependence of
the $(V-I)$ color difference between the BRGB and the MSTO.
In the Figure, $(V-I)_{TO}$ is the $(V-I)$ color of the MSTO and $V_{+0.05}$ is
the $V$ magnitude of the upper MS at a point 0.05 mag redder than the MSTO.
By comparing the color difference at $V - V_{+0.05}$ = $-$2.5 mag,
we obtain $\delta[\Delta(V-I)]$/$\delta$[Fe/H] = $-$0.040 mag dex$^{-1}$,
which is larger than found in the $(B-V)$ color difference
between the BRGB and the MSTO at $V - V_{+0.05}$ = $-$2.5 mag
using the same model isochrones,
$\delta[\Delta(B-V)]$/$\delta$[Fe/H] = $-$0.013 mag dex$^{-1}$
(see the inset of the Figure).
Therefore, the relative age estimate using the color difference
between the BRGB and the MSTO is more vulnerable to metallicity
in $(V-I)$ than in $(B-V)$.
[Note that this method was initially introduced
using $(B-V)$ color differences.]

Figure~7 shows a plot of $(V-I) - (V-I)_{TO}$ versus $V - V_{+0.05}$
for MS/RGB fiducial sequences of NGC~6293 and M92.
Since HST WFPC2 photometry for M92 using the F555W/F814W passbands
does not exist, we adopted the ground-based M92 $VI$ photometry of
Johnson \& Bolte (1997). In the Figure, we used the enhanced $\alpha$-element
model isochrones of Bergbusch \& VandenBerg (2001)
for [Fe/H] = $-$2.01 and [$\alpha$/Fe] = +0.30  from 10 Gyr to 18 Gyr.
As discussed above, NGC~6293 is $\approx$ 0.2 dex more metal-rich than
M92, which results in $\Delta(V-I)$ $\approx$ 0.008 mag in the sense that
the color difference between the BRGB and the MSTO in NGC~6293 is expected
to be 0.008 mag smaller than that in M92 if these two clusters have
the same ages. As shown in the Figure, we also obtain
$\delta[\Delta(V-I)]$/$\delta\tau$ = $-$0.015 mag Gyr$^{-1}$ using
the model isochrones with [Fe/H] = $-$2.01.
A comparison of the fiducial sequences between NGC~6293
and M92 shows excellent agreement from the lower main-sequence to
the RGB within our measurement error and the clusters' fiducial sequences
appear to be almost identical. Therefore, our result shows that
the relative age difference between the two clusters is less than
$\approx$ 0.5 -- 1 Gyr. It should be noted that at
the highest luminosities, the RGB slopes of
the model isochrones brighter than $V - V_{+0.05}$ $\lesssim$ $-$4.0 mag
do not agree with the fiducial sequences of NGC~6293 and M92.

In Figure~8, we show a plot of $(V-I) - (V-I)_{TO}$ versus $V - V_{+0.05}$ for
fiducial sequences of NGC~6541 and M92. The metallicity effect
appears to be visible. The locations of the lower MS of the two clusters
are slightly different, most likely due to the difference in metallicity
(see the difference in the lower main-sequence between
the models with [Fe/H] = $-$2.31 and $-$1.71 in Figure~6).
In the Figure, the color difference between the BRGB and the MSTO
at $V - V_{+0.05}$ $\approx$ $-$2.5 mag in NGC~6541 appears to be
0.005 -- 0.010 mag smaller than that in M92, indicating that
the difference in the relative age between  NGC~6541 and M92 is very small,
$\lesssim$ 0.5 -- 1 Gyr.  However, recall that our fiducial
sequence of the BRGB in NGC~6541 relies on a small number of stars (n $<$ 10)
as shown in Table~\ref{tab5} and, therefore, the uncertainty is large.
In Figure~9 is a plot of $(V-I) - (V-I)_{TO}$ versus $V - V_{+0.05}$
for the fiducial sequences of NGC~6293 and NGC~6541. Again, our results suggest
that NGC~6293 and NGC~6541 essentially have the same ages within
$\approx$ 0.5 -- 1 Gyr.

\subsection{Interstellar Reddening and Distance Modulus}

We explore the interstellar reddenings and distance moduli of the
clusters following the similar scheme described in Lee et al.\ (2001).
We assume that the clusters' MS/RGB fiducial sequences are identical and
we compare the $V_{+0.05}$ magnitude and the $(V-I)_{TO}$ color of the clusters.

In Figure~10, we adopt $\delta V$ = 1.193 mag and $\delta (V-I)$ = 0.496 mag
for M92 and $\delta V$ = $-$4.008 mag and $\delta (V-I)$ = 0.374 mag
for NGC~2419 to match clusters' [$(V-I)_{TO}$, $V_{+0.05}$] points,
in the sense of NGC~6293 minus M92 or NGC~2419.
In Figure~11, we adopt $\delta V$ = 0.096 mag and $\delta (V-I)$ = 0.190 mag
for M92 and $\delta V$ = $-$5.105 mag and $\delta (V-I)$ = 0.068 mag
for NGC~2419, in the sense of NGC~6541 minus M92 or NGC~2419.
From these values, we derive the interstellar reddening and the distance
modulus of NGC~6293 and NGC~6541 with respect to those of M92 or
NGC~2419, assuming $E(B-V)$ = 0.02 and $(m-M)_0$ = 14.60 for M92
(see Lee et al.\ 2001) and $E(B-V)$ = 0.11 and $(m-M)_0$ = 19.54
for NGC~2419 (Harris et al.\ 1997).

Table~\ref{tab6} lists our interstellar reddening and distance modulus estimates
of the clusters. For NGC~6293, we obtain $E(B-V)$ = 0.40 and
$(m-M)_0$ = 14.61 with respect to those of M92, and $E(B-V)$ = 0.40 and
$(m-M)_0$ = 14.64 with respect to those of NGC~2419.
Our reddening estimates for NGC~6293 agree well with that of Harris (1996),
while our distance modulus estimates are $\approx$ 0.1 mag smaller than
that of Harris (see below).
For NGC~6541, we obtain $E(B-V)$ = 0.17 and $(m-M)_0$ = 14.24
with respect to those of M92, and $E(B-V)$ = 0.16 and $(m-M)_0$ = 14.27
with respect to those of NGC~2419. Our reddening estimates for NGC~6541 are
0.03 mag larger than that of Harris (1996), while the distance modulus
estimates are in good agreement.
The 0.03 mag difference in $E(B-V)$ is likely due to the difference
in metallicity of the clusters. As shown in Figures~8 and 11,
the lower MS fiducial sequences of M92 and NGC~2419 show
a slight disagreement with the NGC~6541 lower MS stars most likely due to
the difference in metallicity, suggesting that a correction for
the different metallicity is necessary.
By comparing $(V-I)_{TO}$ and $V_{+0.05}$ of the enhanced $\alpha$-element
model isochrones for [Fe/H] = $-$2.31, $-$2.14, $-$2.01, and $-$1.84,
we derived the change rates in $(V-I)_{TO}$ and $V_{+0.05}$ with metallicity
$\delta (V-I)_{TO}/\delta$[Fe/H] = 0.075 mag dex$^{-1}$ and
$\delta V_{+0.05}/\delta$[Fe/H] = 0.303 mag dex$^{-1}$ at 15 Gyr.
Assuming the metallicity difference between NGC~6293 and M92 is
$\delta$[Fe/H] $\approx$ 0.45 dex, the correction values
become $\delta (V-I)_{TO}$ = 0.034 mag and $\delta V_{+0.05}$ = 0.136 mag,
in the sense that a higher metallicity causes a redder TO color and
a fainter $V_{+0.05}$ magnitude at a given age.
Applying these correction values, we obtain $E(B-V)$ = 0.14 and
$(m-M)_0$ = 14.19 for NGC~6541 with respect to M92, and
$E(B-V)$ = 0.14 and $(m-M)_0$ = 14.22 with respect to NGC~2419.
Our results compare remarkably well with the earlier summaries
provided by
Harris (1996).

Since the interstellar reddening and the distance modulus of M92 are
probably more accurate than those of NGC~2419, due to M92 being
much closer to the Sun and at a higher Galactic latitude, we adopt
M92 as our reference cluster.
Therefore, our final estimates of the interstellar reddening
and the distance modulus for NGC~6293 are $E(B-V)$ = 0.40 mag and
$(m-M)_0$ = 14.61 mag and those for NGC~6541 are $E(B-V)$ = 0.17 mag and
$(m-M)_0$ = 14.24 mag without a metallicity effect correction and
$E(B-V)$ = 0.14 mag and $(m-M)_0$ = 14.19 mag with such a correction.
Schlegel, Finkbeiner, \& Davis (1998) have provided tools to
estimate reddening total reddening along essentially any
line of sight, which should reflect the total reddening to
our two clusters. The results are E($B-V$) = 0.60 mag for NGC~6293
and E($B-V$) = 0.16 mag for NGC~6541. While the latter agrees
well with our derived value, their estimate for NGC~6293
exceeds ours by 50\%.
Dutra \& Bica (2000) noted
that interstellar reddening estimates by Schlegel et al.\ (1998)
for clusters near the Galactic plane appear to be larger than
those based on stellar contents, possibly due to background dust.
Arce \& Goodman (1999) have cautioned their readers that Schlegel
et al.\ (1998) may over-estimate reddening by a factor of 1.3 to 1.5
in regions with E($B-V$) $> 0.2$ or so. Our results appear to
confirm these cautionary remarks about the use of the Schlegel
et al.\ (1998) reddening maps in regions of high reddening.

We also explore the $V_{HB}$ of the clusters. Figure~12 shows the CMDs of
the HB region of NGC~6293 and NGC~6541. Since both clusters have blue tails,
we select HB stars redder than the ``knee" of the BHB by eye
[$(V-I)_0$ $\gtrsim$ 0.1 mag], which are those with $(V-I)$ $\gtrsim$ 0.65 mag
for NGC~6293 and $\gtrsim$ 0.30 mag for NGC~6541.
We then calculate the mean $V$ magnitude of these HB stars and obtain
$V_{HB}$ = 16.34 $\pm$ 0.04 mag (8 stars) for NGC~6293 and
$V_{HB}$ = 15.35 $\pm$ 0.09 mag (2 stars) for NGC~6541.
In the Figure, we show the HB stars used in our calculations
and the $V_{HB}$ magnitudes of the clusters.
Our $V_{HB}$ for NGC~6293 is 0.16 mag brighter than
that of Harris (1996), who obtained $V_{HB}$ = 16.50 mag using the data
obtained by Janes \& Heasley (1991).  It is thought that this difference
in the NGC~6293 $V_{HB}$ magnitude may explain a 0.1 mag difference
in the distance modulus of NGC~6293 in Table~\ref{tab6}, since the previous distance
modulus of NGC~6293 mainly relies on the $V_{HB}$ magnitude.
For NGC~6541, our $V_{HB}$ is $\approx$ 0.25 mag and 0.15 mag fainter
than those of Alcaino (1979) and Alcaino et al.\ (1997), who obtained
$V_{HB}$ = 15.1 mag and 15.2 mag, respectively.
It should be noted that, however, our $V_{HB}$ for NGC~6541 is rather
uncertain since our measurement is based on only two HB stars.
It also should be noted that the HB stars that we used in our calculations
appear to be located in the instability strip\footnote{In Figure~12,
we also show the first harmonic blue edge and the fundamental red edge
of M3 RR Lyrae variables (Carretta et al.\ 1998).}
and they are likely either membership RR Lyrae variables or
non-variable off-cluster field populations that appear to lie inside
the instability strip.
As shown in Table~\ref{tab1}, the HB type ($B-R/B+V+R$) of NGC~6541 is 1.00,
i.e.\ NGC~6541 is known to have neither RR Lyrae variables nor red HB stars,
and therefore the HB stars in Figure~12 could be off-cluster field populations.
Since our photometric measurements of bright stars in our fields rely on
the three short exposure frames (see Table~\ref{tab2}), our $V_{HB}$
estimate may not be appropriate to estimate mean magnitude of RR Lyrae
variables.

If we adopt the intensity-weighted $\langle M_{V}(HB) \rangle$ =
$\langle M_{V}(RR) \rangle$
= 0.44 $\pm$ 0.03 mag for M92 (see Lee et al.\ 2001), $(m-M)_0$ = 14.66
$\pm$ 0.05 mag for NGC~6293 and $(m-M)_0$ = 14.38 $\pm$ 0.09 mag for
NGC~6541 if $E(B-V)$ = 0.17 and $(m-M)_0$ = 14.47 $\pm$ 0.09 mag
if $E(B-V)$ = 0.14.
Our distance modulus using the HB stars agrees well with that
in Table~\ref{tab6} for NGC~6293. For NGC~6541,\footnote{ The luminosity
of the RR Lyrae variables or HB stars are also affected by metallicity.
If we use the relation of Carney, Storm, \& Jones (1992),
\begin{equation}
\langle M_{V}(RR) \rangle = 0.16 \mathrm{[Fe/H]} + 1.02,
\end{equation}
$\langle M_{V}(RR) \rangle$ for NGC~6541 is $\approx$ 0.06 mag fainter
than that of M92 and consequently $(m-M)_0$ = 14.41, which is still larger
than those in Table~\ref{tab6}.}
the distance modulus using the two HB stars is $\approx$ 0.15 mag larger
than that in Table~\ref{tab6}, suggesting that perhaps $V_{HB}$ $\approx$ 15.2 mag
is correct for NGC~6541, which is adopted by Harris (1996) and
Alcaino et al.\ (1997).
If we adopt our distance modulus values $(m-M)_0$ = 14.61 mag and 14.24 mag
for NGC~6293 and NGC~6541, the distances of the clusters from
the Sun become 8.4 kpc and 6.9 kpc, respectively.
The Galactocentric distances of NGC~6293 and NGC~6541 are 1.2 kpc and 2.3 kpc,
respectively, for $R_0$ = 8.0 kpc (Reid 1993). Therefore, NGC~6293 and
NGC~6541 are clearly located in the Galaxy's central regions.

\subsection{Differential Reddening in NGC~6293?}

In Figure~10, the MS/RGB fiducial sequences of M92 and NGC~2419 show
an excellent agreement with NGC~6293 MS/RGB stars,
while the HB sequences of M92 and NGC~2419 appear to be slightly brighter
and bluer than the NGC~6293 HB stars.
On the other hand, the NGC~6541 BHB stars show better agreement with
the HB fiducial sequences of M92 or NGC~2419 in Figure~11.

The disagreement in HB magnitudes or colors
between NGC~6293 and M92 (or NGC~2419) seems hard to explain, since
NGC~6293 and M92 appear to have similar chemical compositions and ages
as we discussed above. We can imagine two possible explanations,
predicated on the assumption that the CMDs for all three clusters
should be very similar. First, there is apparently a source temperature
sensitivity to reddening corrections.
This can be a significant
effect in the HST WFPC2 F550W/F814W photometric system, in the sense that
the interstellar extinction values for NGC~6293 HB stars are larger than
those for NGC~6293 MS/RGB stars due to differences surface temperature.
As Holtzman et al.\ (1995a) pointed out, the bandwidths of
HST WFPC2 F555W/F814 filter system are wider
than the ground-based filter system, so the interstellar extinction
in the HST WFPC2 photometric system is slightly vulnerable to
the temperature (i.\ e., $F_{\lambda}$) of the source
[for example, see the Appendix~A of Lee et al.\ (2002) for
the detailed discussion of the temperature dependence in the interstellar
extinction law for the HST NICMOS photometric system].
More recently, Romaniello et al.\ (2002) presented interstellar extinction
coefficients of the HST WFPC2 photometric system for selected stellar
temperatures (see their Table~8). Using their results, we obtain
differential reddening correction values between $T_{eff}$ = 6500 K
($\approx$ $T_{eff}$ of the MSTO) and 10000 K ($\approx$ $T_{eff}$ of the HB),
of $\Delta V$ = 0.027 mag and $\Delta (V-I)$ = 0.018 mag
assuming $E(B-V)$ = 0.40 for NGC~6293,
in the sense that blue HB stars suffer more extinction than MSTO stars.
We show our results in Figure 13. In the upper panel of the Figure,
we show the differential reddening correction vector
due to the difference in temperature by an arrow.
In the bottom panel of the Figure, we show the NGC~6293 CMD after applying
the differential reddening correction due to the temperature difference
between the MSTO and the HB. The agreement in HB locations is slightly
improved in the Figure.

However, many of the NGC~6293 HB stars still appear to have fainter magnitudes
or redder colors than  M92 HB stars,
suggesting that the discrepancy may not be mainly due to
the {\em induced} differential reddening due to differences in the temperatures
of the stars. {\em True} differential reddening and variations in
interstellar absorption may also be at work. We approach this
possibility by considering
the apparently under-luminous HB stars in the field NGC~6293-F1
by eye (see Figure~14). We show their spatial distribution in Figure~15.
In the Figure, gray dots represent the HB stars with normal luminosity
and black dots represent the under-luminous HB stars in the field NGC~6293-F1.
Figure~15 clearly indicates that most under-luminous HB stars
are located to the north of the cluster (the upper left corner of the Figure),
suggesting that differential reddening exists in the cluster.
We also explore this differential reddening effect using the NGC~6293 RGB stars.
We select stars with $X$ $\leq$ 300 and $Y$ $\geq$ 400
detected in the field NGC~6293-F1 (inside the dotted lines in Figure~15;
this area does not overlap with the field NGC~6293-F2) and we show their CMD
superposed on the CMD of stars detected in the rest of the field in Figure~16.
As can be seen, the colors of RGB stars in this region appears to be
$\Delta(V-I)$ $\approx$ 0.03 -- 0.05 mag redder than the mean color of
the fiducial sequence of the cluster.
Therefore, if this color difference is due to the differential reddening,
the interstellar extinction toward the north of the cluster is
$\Delta E(B-V)$ = 0.02 -- 0.04 mag and $\Delta V$ = 0.06 -- 0.12 mag larger.
In Figure~17, we show the CMD of the HB region in the field NGC~6293-F1
after applying the mean differential reddening correction,
$\Delta(V-I)$ = 0.04 mag and  $\Delta V$ = 0.08 mag.
It should be recalled that this differential reddening correction value
came from the RGB stars and applying this correction value to
the NGC~6293 HB stars shows good agreement with the M92 HB fiducial
sequence.

We conclude that the disagreement in HB magnitudes
or colors between NGC~6293 and M92 (or NGC~2419) in Figure~10 are mainly due to
true differential reddening across NGC~6293.

\section{AGES OF THE METAL-POOR INNER HALO CLUSTERS}

The absolute and relative ages of globular clusters are of considerable
interest in trying to resolve {\em when} star formation began in
the Galaxy and {\em how long} the Galactic halo formed stars. In
this study, we are concerned primarily with the most metal-poor clusters,
which are presumably the oldest, and the key question to which we seek
the answer is whether star formation began at different times in the
higher density central regions compared to the lower density outer
regions. We are not the first to address this question, of course.
Rosenberg et al.\ (1999) employed two relative age-dating techniques,
one relying on the magnitude difference between the horizontal branch
and the main sequence turn-off (the ``vertical" method) and the
other relying on the color difference between the turn-off and a
point 2.5 mag brighter on the red giant branch (the ``horizontal"
method). They found that the most metal-poor clusters had ages
that were not distinguishably different, but their innermost
metal-poor cluster ([Fe/H] $\leq\ -1.7$) lies at $R_{\rm GC}$ = 6.0 kpc.
Salaris \& Weiss (2002) also addressed globular cluster relative ages,
relying on a mix of ``vertical" and ``horizontal" methods, although
in the latter case they relied on the color differences defined
by VandenBerg et al.\ (1990), which is the same approach that we
have adopted. They found results very similar to Rosenberg et al.\ (1999),
which is not too surprising since very similar clusters were analyzed
using very similar techniques. Again, the innermost metal-poor cluster,
NGC~6397, lies at $R_{\rm GC}$ = 6.0 kpc. The most recent such study
is that of De~Angeli et al.\ (2005), who relied on a mix of HST and
ground-based data of high quality. Using the Zinn \& West (1984)
metallicity scale, their sample includes four clusters with
$R_{\rm GC}$ $<$ 5.0 kpc: NGC~6093 (3.8 kpc); NGC~6273 (1.6 kpc);
NGC~6287 (1.7 kpc); and NGC 6809 (3.9 kpc). Using a ``normalized
age", defined to be the ratio of the derived age to that of the
ensemble of all metal-poor clusters, these four clusters were
found to have normalized ages of $0.97 \pm 0.07$, $0.96 \pm 0.03$,
$1.05 \pm 0.11$, and $1.05 \pm 0.12$. In other words, not only
do the ensemble of metal-poor clusters show no detectable differences
in derived relative ages, neither do the innermost metal-poor clusters.

Our work on relative cluster ages has concentrated on the innermost
globular clusters {\em with blue horizontal branches}. Our logic,
like that of van den Bergh (1993), has
been that if the color of the horizontal branch is determined by
both metallicity and age, these may be the oldest clusters in the
Galaxy. Our earlier work on NGC~6287
(Lee et al.\ 2001) showed that the most metal-poor inner halo globular cluster
NGC~6287 and M92 have the same ages within $\pm$2 Gyr ($\approx$ 14 \%
if the absolute age of M92 is 14 Gyr; VandenBerg 2000). De~Angeli et al.\ (2005)
likewise concluded that NGC~6287 was not distinguishably different
in age from outer halo metal-poor clusters, presumably using the
same HST data as did we.

In this paper we have discussed the relative ages of the inner halo globular clusters
NGC~6293 and NGC~6541 with respect to the intermediate halo globular
cluster M92 and the remote halo globular cluster NGC~2419. The blue
horizontal branches that dominate these clusters have made employment
of any ``vertical" method of relative age dating problematical, and
we have had to rely, instead, on the ``horizontal" method defined
by VandenBerg et al.\ (1990).
We have shown here that NGC~6293 and NGC~6541 essentially have
the same ages as M92 and NGC~2419 within $\pm$1 Gyr ($\approx$ 7 \%),
in spite of very different current Galactocentric distances.
We have also discussed the effect of $\alpha$-elements on our relative age
estimates and we suggested that their contribution is negligibly small,
in particular for NGC~6287 and NGC~6293.
Derived ages of globular clusters also depend on the adopted helium
abundances, since the helium abundance governs stellar evolutionary timescales
and stellar internal structures (primarily due to its influence on
mean molecular weight).
The most frequently used method for the helium abundance estimation of
the globular clusters is the $R$-method, where $R$ is defined to be
the number ratio of HB stars and RGB stars brighter than the zero-age HB stars,
$N_{HB}/N_{RGB}$, (Buzzoni et al.\ 1983; Caputo et al.\ 1987).
The number ratio $R$ is then related to the helium abundance of the clusters.
In our study, we were not able to estimate the helium abundances of our program
clusters using the $R$-method mainly due to the saturation of the bright
RGB stars. Also we do not have photometric measurements for the off-cluster
field populations to correct the field star contamination in deriving
the number ratio. Therefore, the helium abundance of our program clusters
is still an open question. However, the
recent study of Sandquist (2000) suggested that
the helium abundance of the globular clusters appears
to be constant (Y $\approx$ 0.20) with metallicity, as expected if $\Delta$Y
scales proportionately with $\Delta$Z.

In Figure~18, we show relative ages of the globular clusters as functions
of metallicity and the Galactocentric distance using the combined results
of our work, Harris et al.\ (1997), Heasley et al.\ (2000), Lee et al.\ (2001),
Rosenberg et al.\ (1999),\footnote{It should be noted that the relative
age estimates of Rosenberg (1999) may be slightly inaccurate.
Lee \& Carney (1999b) argued that M2, which is most likely as old as M92,
is $\approx$ 2 Gyr older than M3, while Rosenberg et al.\ (1999) claimed that
M3 and M92 have the same ages. VandenBerg (2000) also claimed that M3 appears
to be $\approx$ 1 - 1.5 Gyr younger than M92.
Recently, Rey et al.\ (2001) studied M3 vs M13 and they concluded that
M13 is 1.7 $\pm$ 0.7 Gyr older than M3, while Rosenberg et al.\ (1999)
suggested the same ages between the two clusters.
Another example is M5 compared to NGC~6752. VandenBerg (2000) obtained
$\approx$ 2 Gyr younger ages than did  Rosenberg et al.\ (1999) for
these clusters. Although Rosenberg et al.\ (1999) claimed that their results
are based on the homogenous data sets, an independent investigation using deep
main-sequence photometry  of their sample would be desirable.
Figure~18 relies mainly on Rosenberg et al.\ (1999).}
and Stetson et al.\ (1999). The Figure shows that the ages of the oldest
globular clusters (mostly the old halo population)  in our Galaxy
do not vary with metallicity ([Fe/H] $\leq$ $-$1.0) or Galactocentric distance.
In particular, our work clearly shows that there is no age gradient
in the inner part of our Galaxy. Thus, our results are consistent with
the idea that the globular cluster formation must have been triggered
almost everywhere at about the same time in our Galaxy. Since Harris et al. (1997) claimed
that the very remote halo cluster NGC~2419 ($R_{GC}$ $\approx$ 90 kpc) and M92
essentially have the same age, our result and that of Lee et al.\ (2001)
extend this to the three most metal-poor inner halo ($R_{GC}$ $<$ 3 kpc)
globular clusters in our Galaxy.

\section{CONCLUSIONS}

We have presented HST PC2 photometry for the inner halo globular clusters
NGC~6293 and NGC~6541. Our CMD for NGC~6293 shows a strong BHB population,
consistent with its low metallicity and old age.
It also appears to have blue straggler stars and
a future study related to these objects is desirable.
We could not investigate the RGB-tip luminosity, which was claimed by
Janes \& Heasley (1991) to be abnormally faint,
since the bright RGB stars were saturated in our images.
Our magnitudes and colors for the NGC~6541 main-sequence TO are in good
agreement with the ground-based $VI$ photometry of Alcaino (1997).
Our CMD for NGC~6541 appears to be deficient in bright RGB stars,
confirming the finding of Alcaino (1997).

We have discussed the interstellar reddening and the distance modulus
of NGC~6293 and NGC~6541 with respect to those of M92.
For NGC~6293, our interstellar reddening estimate is consistent with
previous results, while our distance modulus is $\approx$ 0.1 mag
smaller than the previous estimate by Harris (1996).
We also discussed the differential reddening across NGC~6293. It appears
that the interstellar reddening value of NGC~6293 varies by
$\Delta E(B-V)$ $\approx$ 0.02 -- 0.04 mag.
For NGC~6541, our age-dating method, which makes use of M92 as a template,
appears to suffer from modest metallicity difference effects.
Our interstellar reddening and distance modulus of NGC~6541 are
$E(B-V)$ = 0.17 and $(m-M)_0$ = 14.24 without the correction for
the metallicity effect, and $E(B-V)$ = 0.14 and $(m-M)_0$ = 14.19
with such a correction. Nevertheless, NGC~6293 and NGC~6541 are clearly located
in the Galaxy's central regions ($R_{GC} \leq$ 3 kpc).

The most interesting result of our study is that the inner halo clusters
NGC~6293 and NGC~6541 essentially have ages that are indistinguishably
different from one of the oldest globular clusters in our Galaxy M92,
consistent with the previous result of NGC~6287 by Lee et al.\ (2001),
and the large study of De~Angeli et al.\ (2005).
Furthermore, they appear to have the same ages as the most remote
metal-poor globular cluster NGC~2419 ($R_{GC} \approx$ 90 kpc).
Our results strongly support the idea that the globular cluster formation
must have begun everywhere at the same time to within $\approx$ 0.5 -- 1 Gyr
in our Galaxy.

\acknowledgments
This research was supported by the National Aeronautics and Space
Administration (NASA) grant number GO-07318.04-96A from the Space Telescope
Science Institute, which is operated by the Association of Universities
for Research in Astronomy (AURA), Inc., under NASA contract NAS 5-26555 and
the National Science Foundation grants AST-9619381, AST-9988156,
and AST-0305431 to the University of North Carolina.
Support for this work was also provided in part by the Korea Science
and Engineering Foundation (KOSEF) to the Astrophysical Research Center
for the Structure and Evolution of the Cosmos (ARCSEC).


\clearpage

\begin{deluxetable}{ccccccccc}
\tablecaption{Cluster Properties.}
\tablenum{1}
\tablewidth{0pc}
\tablehead{
\colhead{ID} & \colhead{[Fe/H]\tablenotemark{a}} &
\colhead{$E(B-V)$} & \colhead{{\it l}} & \colhead{{\it b}} &
\colhead{$R_{GC}$} & \colhead{HB type\tablenotemark{b}} &
\colhead{$c$\tablenotemark{c}} & \colhead{$v_r$\tablenotemark{a}}\\
\colhead{} & \colhead{} & \colhead{} & \colhead{} & \colhead{} &
\colhead{$(kpc)$} & \colhead{} & \colhead{} & \colhead{(km/s)}}
\startdata
NGC~6293 & -1.99 & 0.41 & 357.62 &  ~~~7.83 & 1.4 & 0.90 & 2.5 & $-$159.9 \\
NGC~6541 & -1.76 & 0.14 & 349.29 & $-$11.18 & 2.2 & 1.00 & 2.0 & $-$167.5 \\
\enddata
\tablenotetext{a}{Based on the high-resolution echelle spectroscopy of
the two giants (Lee \& Carney, 2002).}
\tablenotetext{b}{HB type = $B-R/B+V+R$ (Lee et al.\ 1994).}
\tablenotetext{c}{Cluster central concentration, $c = \log r_t/r_c$}
\label{tab1}
\end{deluxetable}

\clearpage

\begin{deluxetable}{ccccc}
\tablecaption{Journal of Observations for HST PC2 photometry.}
\tablenum{2}
\tablewidth{0pc}
\tablehead{
\colhead{ID} & \colhead{Field} & \colhead{Passband} &
\colhead{Date/Time} & \colhead{$t_{exp}$} \\
\colhead{} & \colhead{} & \colhead{} &
\colhead{(UT start)} & \colhead{(sec)} }
\startdata
NGC~6293 & F1 & F555W & 1995 Mar 23 12:44 & ~40.0 $\times$ 3 \\
         &    & F555W & 1995 Mar 23 12:50 & 500.0 $\times$ 4 \\
         &    & F814W & 1995 Mar 23 14:23 & ~60.0 $\times$ 3 \\
         &    & F814W & 1995 Mar 23 14.32 & 700.0 $\times$ 4 \\
         & F2 & F555W & 1995 Mar 23 17:24 & ~40.0 $\times$ 3 \\
         &    & F555W & 1995 Mar 23 17:30 & 500.0 $\times$ 4 \\
         &    & F814W & 1995 Mar 23 19:12 & ~60.0 $\times$ 3 \\
         &    & F814W & 1995 Mar 23 19:21 & 700.0 $\times$ 4 \\
  &  &  & & \\
NGC~6541 & F1 & F555W & 1994 Oct 15 00:08 & ~12.0 $\times$ 3 \\
         &    & F555W & 1994 Oct 15 00:14 & 140.0 $\times$ 4 \\
         &    & F814W & 1994 Oct 15 00:33 & ~20.0 $\times$ 3 \\
         &    & F814W & 1994 Oct 15 00:39 & 260.0 $\times$ 3 \\
         & F2 & F555W & 1994 Oct 15 01:52 & ~12.0 $\times$ 3 \\
         &    & F555W & 1994 Oct 15 01:58 & 140.0 $\times$ 4 \\
         &    & F814W & 1994 Oct 15 02:17 & ~20.0 $\times$ 3 \\
         &    & F814W & 1994 Oct 15 03:13 & 260.0 $\times$ 4 \\
\enddata
\label{tab2}
\end{deluxetable}

\clearpage
\begin{deluxetable}{crrrrrr}
\tablecaption{Color-magnitude diagram data for NGC~6293.}
\tablenum{3}
\tablewidth{0pt}
\tablehead{
\colhead{Field\tablenotemark{1}} & \colhead{~~ID} &
\colhead{X\tablenotemark{2}} & \colhead{Y\tablenotemark{2}} &
\colhead{$V$} & \colhead{$I$} & \colhead{$V-I$}}
\startdata
1 & 154 & 189.826 & 60.580 & 17.648 & 16.194 & 1.454 \\
1 & 194 & 652.373 & 61.169 & 18.108 & 16.689 & 1.419 \\
1 & 546 & 113.289 & 62.445 & 19.499 & 18.197 & 1.302 \\
1 & 585 & 371.216 & 67.719 & 19.493 & 18.397 & 1.096 \\
1 & 435 & 292.458 & 69.570 & 19.231 & 17.999 & 1.232 \\
1 & 506 &  72.605 & 71.835 & 19.384 & 18.115 & 1.269 \\
1 & 250 & 660.657 & 72.716 & 18.361 & 17.036 & 1.324 \\
1 & 298 & 303.934 & 73.579 & 18.760 & 17.473 & 1.286 \\
1 & 543 & 339.264 & 78.241 & 19.455 & 18.309 & 1.145 \\
\enddata
\tablenotetext{1}{``1" -- F1 only; ``2" -- F2 only; ``o" -- overlap region}
\tablenotetext{2}{The pixel coordinate, where the image scale is
0.0455 arcsec/pixel}
\label{tab3}
\end{deluxetable}

\clearpage

\begin{deluxetable}{ccccrcccccr}
\tablecaption{Fiducial sequence for NGC~6293.}
\tablenum{4}
\tablewidth{0pc}
\tablehead{
\colhead{$V$} & \colhead{$\pm$} &
\colhead{$V-I$} & \colhead{$\pm$} & \colhead{n} &
\colhead{} &
\colhead{$V$} & \colhead{$\pm$} &
\colhead{$V-I$} & \colhead{$\pm$} & \colhead{n}}
\startdata
15.574 &0.020 & 1.509 &0.010 &  4 && 20.198 &0.003 & 1.064 &0.001 &415  \\
15.783 &0.025 & 1.486 &0.008 &  5 && 20.402 &0.003 & 1.076 &0.002 &392  \\
16.005 &0.020 & 1.454 &0.008 & 10 && 20.604 &0.003 & 1.093 &0.002 &471  \\
16.202 &0.025 & 1.431 &0.007 &  5 && 20.797 &0.003 & 1.108 &0.002 &446  \\
16.440 &0.019 & 1.418 &0.005 &  5 && 20.997 &0.003 & 1.126 &0.002 &505  \\
16.591 &0.014 & 1.399 &0.007 &  6 && 21.200 &0.003 & 1.151 &0.002 &493  \\
16.749 &0.016 & 1.390 &0.006 &  6 && 21.394 &0.003 & 1.175 &0.002 &482  \\
16.988 &0.014 & 1.381 &0.005 & 20 && 21.601 &0.003 & 1.206 &0.002 &489  \\
17.214 &0.011 & 1.358 &0.006 & 10 && 21.799 &0.003 & 1.228 &0.002 &507  \\
17.414 &0.014 & 1.357 &0.004 & 17 && 21.999 &0.003 & 1.264 &0.002 &435  \\
17.611 &0.012 & 1.345 &0.005 & 20 && 22.198 &0.003 & 1.296 &0.002 &422  \\
17.807 &0.014 & 1.329 &0.005 & 25 && 22.400 &0.003 & 1.327 &0.002 &390  \\
18.007 &0.012 & 1.327 &0.006 & 23 && 22.606 &0.003 & 1.368 &0.003 &316  \\
18.189 &0.009 & 1.313 &0.004 & 33 && 22.800 &0.003 & 1.414 &0.002 &307  \\
18.416 &0.008 & 1.301 &0.003 & 38 && 23.000 &0.003 & 1.454 &0.003 &302  \\
18.611 &0.010 & 1.283 &0.004 & 38 && 23.203 &0.004 & 1.489 &0.003 &244  \\
18.797 &0.008 & 1.272 &0.005 & 39 && 23.393 &0.004 & 1.535 &0.003 &179  \\
19.011 &0.007 & 1.238 &0.005 & 56 && 23.604 &0.004 & 1.583 &0.004 &168  \\
19.215 &0.006 & 1.177 &0.004 & 76 && 23.798 &0.005 & 1.632 &0.004 &134  \\
19.419 &0.004 & 1.110 &0.003 &186 && 23.996 &0.005 & 1.666 &0.004 &143  \\
19.607 &0.003 & 1.062 &0.002 &245 && 24.190 &0.005 & 1.715 &0.005 &105  \\
19.801 &0.003 & 1.052 &0.002 &291 && 24.400 &0.006 & 1.771 &0.006 & 83  \\
20.009 &0.003 & 1.051 &0.002 &332 && 24.589 &0.007 & 1.803 &0.006 & 65  \\
\enddata
\label{tab4}
\end{deluxetable}

\clearpage

\begin{deluxetable}{ccccrcccccr}
\tablecaption{Fiducial sequence for NGC~6541.}
\tablenum{5}
\tablewidth{0pc}
\tablehead{
\colhead{$V$} & \colhead{$\pm$} &
\colhead{$V-I$} & \colhead{$\pm$} & \colhead{n} &
\colhead{} &
\colhead{$V$} & \colhead{$\pm$} &
\colhead{$V-I$} & \colhead{$\pm$} & \colhead{n}}
\startdata
16.844 & 0.014 & 1.009 & 0.010 &   3 && 21.209 & 0.006 & 1.031 & 0.003 &  84 \\
17.227 & 0.000 & 0.983 & 0.000 &   1 && 21.394 & 0.006 & 1.073 & 0.004 &  93 \\
17.428 & 0.025 & 0.976 & 0.005 &   6 && 21.607 & 0.006 & 1.123 & 0.004 & 106 \\
17.590 & 0.032 & 0.965 & 0.006 &   5 && 21.798 & 0.006 & 1.165 & 0.004 & 105 \\
17.842 & 0.022 & 0.950 & 0.006 &   8 && 21.993 & 0.005 & 1.207 & 0.004 & 114 \\
18.017 & 0.018 & 0.894 & 0.005 &  13 && 22.189 & 0.006 & 1.251 & 0.004 &  91 \\
18.171 & 0.007 & 0.850 & 0.008 &   2 && 22.396 & 0.006 & 1.312 & 0.004 &  97 \\
18.402 & 0.016 & 0.772 & 0.007 &  19 && 22.598 & 0.006 & 1.368 & 0.005 &  95 \\
18.589 & 0.011 & 0.756 & 0.004 &  29 && 22.802 & 0.006 & 1.424 & 0.004 &  96 \\
18.811 & 0.009 & 0.745 & 0.004 &  39 && 23.001 & 0.006 & 1.482 & 0.005 &  90 \\
19.023 & 0.008 & 0.750 & 0.004 &  40 && 23.203 & 0.006 & 1.547 & 0.005 & 103 \\
19.196 & 0.008 & 0.765 & 0.003 &  64 && 23.404 & 0.006 & 1.564 & 0.006 &  89 \\
19.395 & 0.007 & 0.778 & 0.003 &  73 && 23.598 & 0.006 & 1.598 & 0.005 & 102 \\
19.613 & 0.007 & 0.798 & 0.003 &  65 && 23.803 & 0.007 & 1.649 & 0.007 &  72 \\
19.815 & 0.007 & 0.822 & 0.004 &  66 && 24.003 & 0.007 & 1.709 & 0.006 &  76 \\
19.995 & 0.007 & 0.832 & 0.003 &  78 && 24.201 & 0.007 & 1.770 & 0.006 &  78 \\
20.213 & 0.006 & 0.863 & 0.003 &  96 && 24.396 & 0.008 & 1.784 & 0.008 &  62 \\
20.392 & 0.005 & 0.892 & 0.003 & 107 && 24.589 & 0.008 & 1.873 & 0.006 &  51 \\
20.596 & 0.005 & 0.928 & 0.003 & 100 && 24.811 & 0.009 & 1.962 & 0.009 &  33 \\
20.802 & 0.005 & 0.953 & 0.003 & 101 && 24.996 & 0.013 & 1.961 & 0.011 &  20 \\
20.998 & 0.006 & 0.993 & 0.003 & 102 &&        &       &       &       &     \\
\enddata
\label{tab5}
\end{deluxetable}

\clearpage

\begin{deluxetable}{ccccccccc}
\tablecaption{Interstellar reddening and distance modulus for NGC~6293
and NGC~6541.}
\tablenum{6}
\tablewidth{0pc}
\tablehead{
\multicolumn{2}{c}{NGC~6293} & \multicolumn{1}{c}{} &
\multicolumn{4}{c}{NGC~6541} & \multicolumn{1}{c}{} &
\multicolumn{1}{c}{Note} \\
\cline{1-2}\cline{4-7} \\
\colhead{$E(B-V)$} & \colhead{$(m-M)_0$} & \colhead{} &
\colhead{$E(B-V)$} & \colhead{$(m-M)_0$} &
\colhead{$E(B-V)$\tablenotemark{a}} & \colhead{$(m-M)_0$\tablenotemark{a}} &
\colhead{} & \colhead{}}
\startdata
0.40 & 14.61 & & 0.17 & 14.24 & 0.14 & 14.19 && (1) \\
0.40 & 14.64 & & 0.16 & 14.27 & 0.14 & 14.22 && (2) \\
0.41 & 14.72 & & 0.14 & 14.24 &      &       && (3) \\
\enddata
\tablenotetext{a}{Corrected for the metallicity effect.}
\tablecomments{
(1) With respect to M92;
(2) with respect to NGC~2419;
(3) Harris (1996)}
\label{tab6}
\end{deluxetable}

\clearpage

\begin{figure}
\epsscale{1}
\figurenum{1}
\plotone{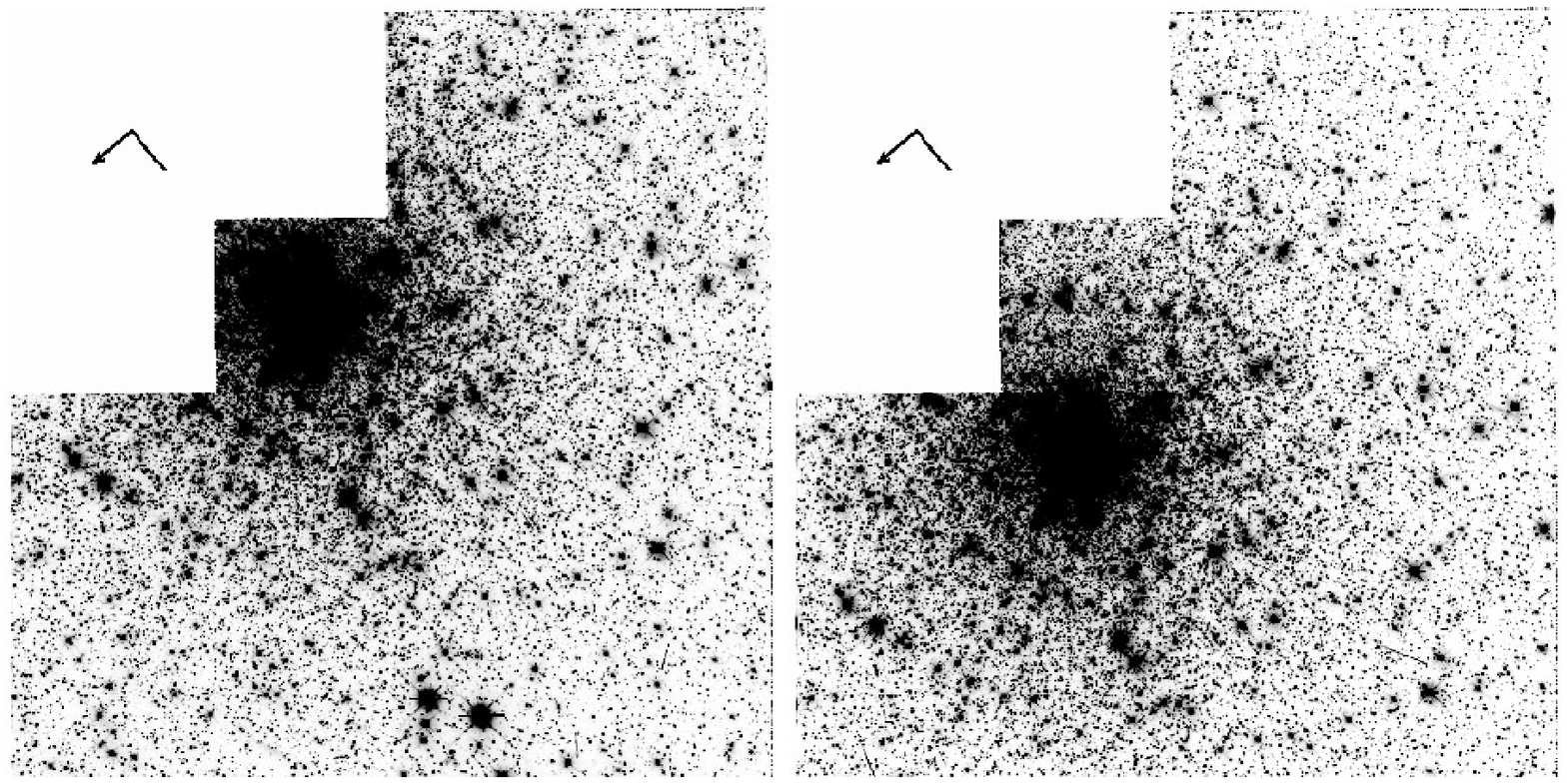}
\caption
{Final F555W images of fields NGC~6293-F1 and F2. North and east
are indicated.}
\end{figure}

\begin{figure}
\epsscale{1}
\figurenum{2}
\plotone{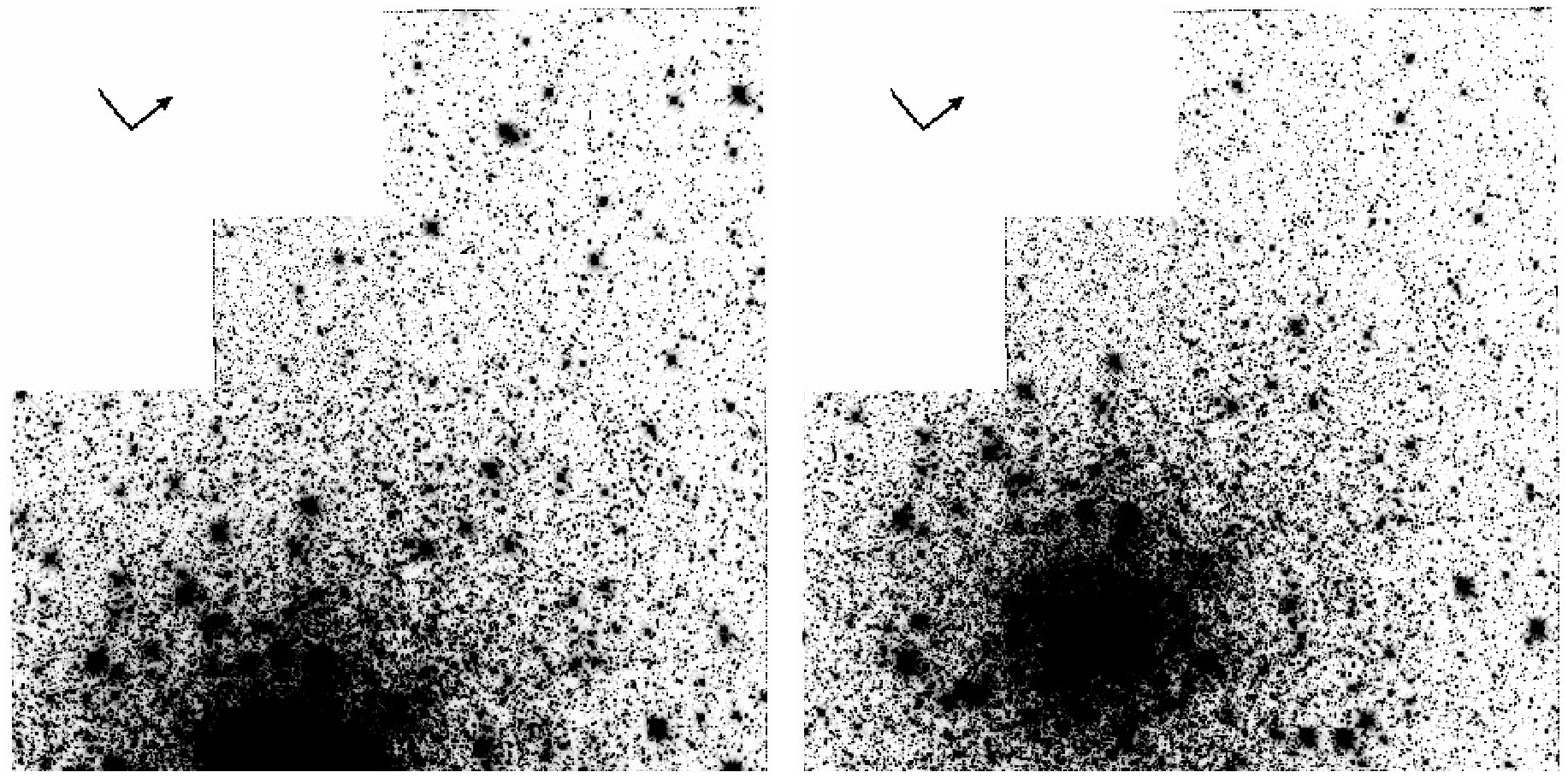}
\caption
{Final F555W images of fields NGC~6541-F1 and F2. North and east
are indicated.}
\end{figure}

\begin{figure}
\epsscale{1}
\figurenum{3}
\plotone{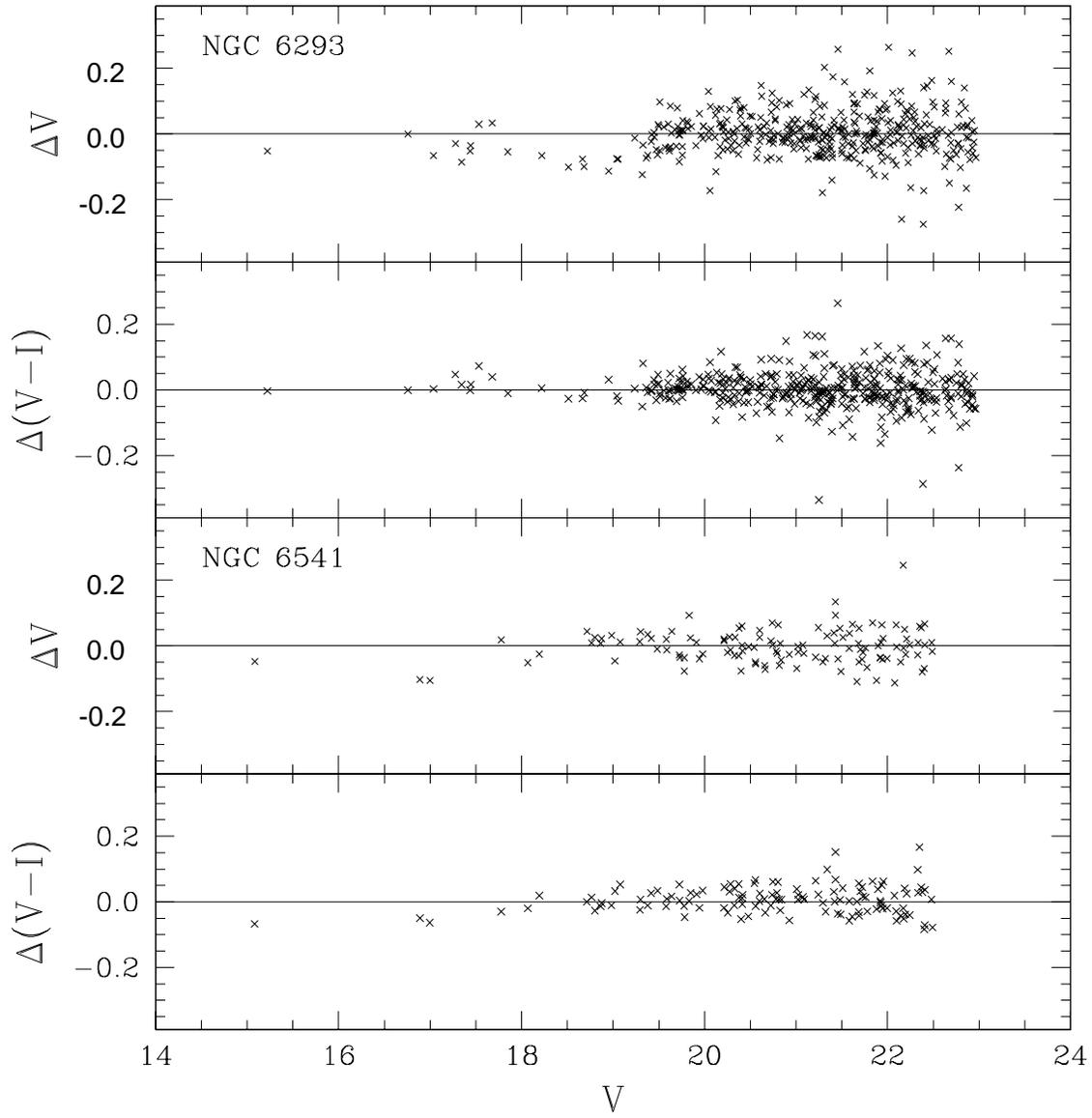}
\caption
{Comparisons of overlapped regions in NGC~6293 and NGC~6541.}
\end{figure}

\begin{figure}
\epsscale{1}
\figurenum{4}
\plotone{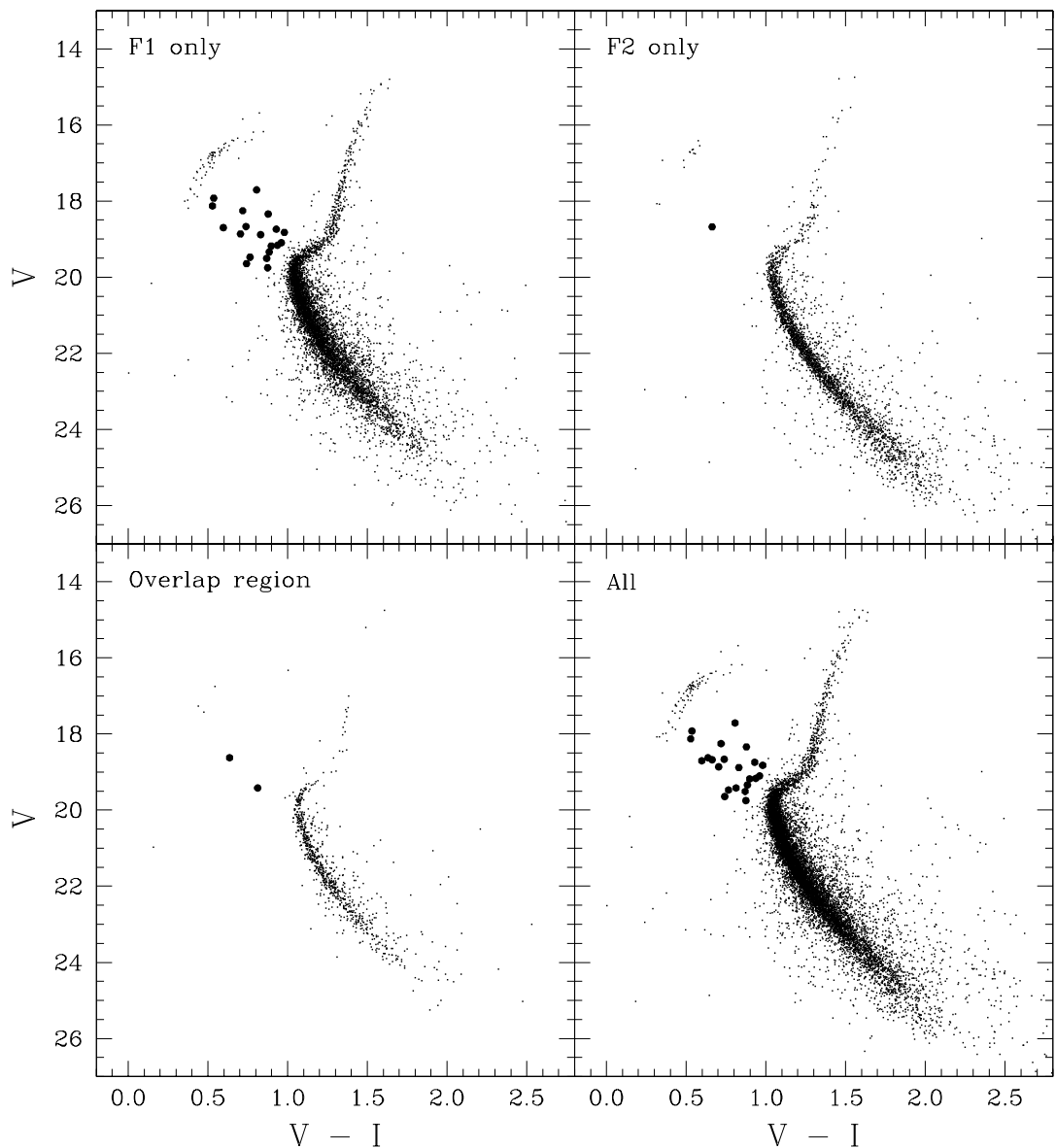}
\caption
{CMDs of NGC~6293 for the stars in the fields F1 (6761 stars),
F2 (3150 stars), overlap region (844 stars), and for the combined data.
Newly found 22 BSS candidates (19 in the field F1 only,
1 in the field F2 only, and 2 in the overlap region)
are represented by filled circles.}
\end{figure}

\begin{figure}
\epsscale{1}
\figurenum{5}
\plotone{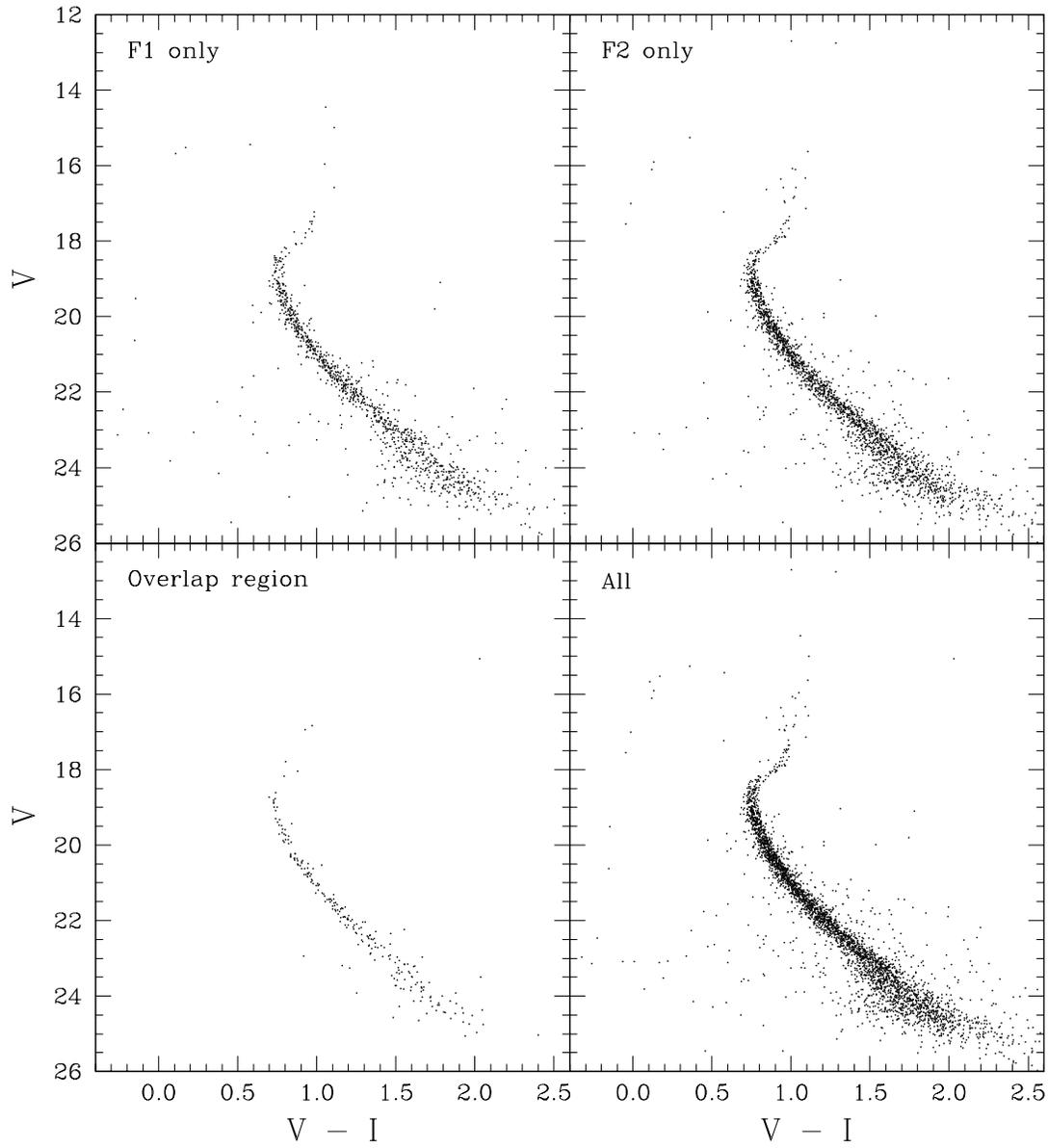}
\caption
{CMDs of NGC~6541 for the stars in the fields F1 (1086 stars),
F2 (2162 stars), overlap region (235 stars), and for the combined data.}
\end{figure}

\begin{figure}
\epsscale{1}
\figurenum{6}
\plotone{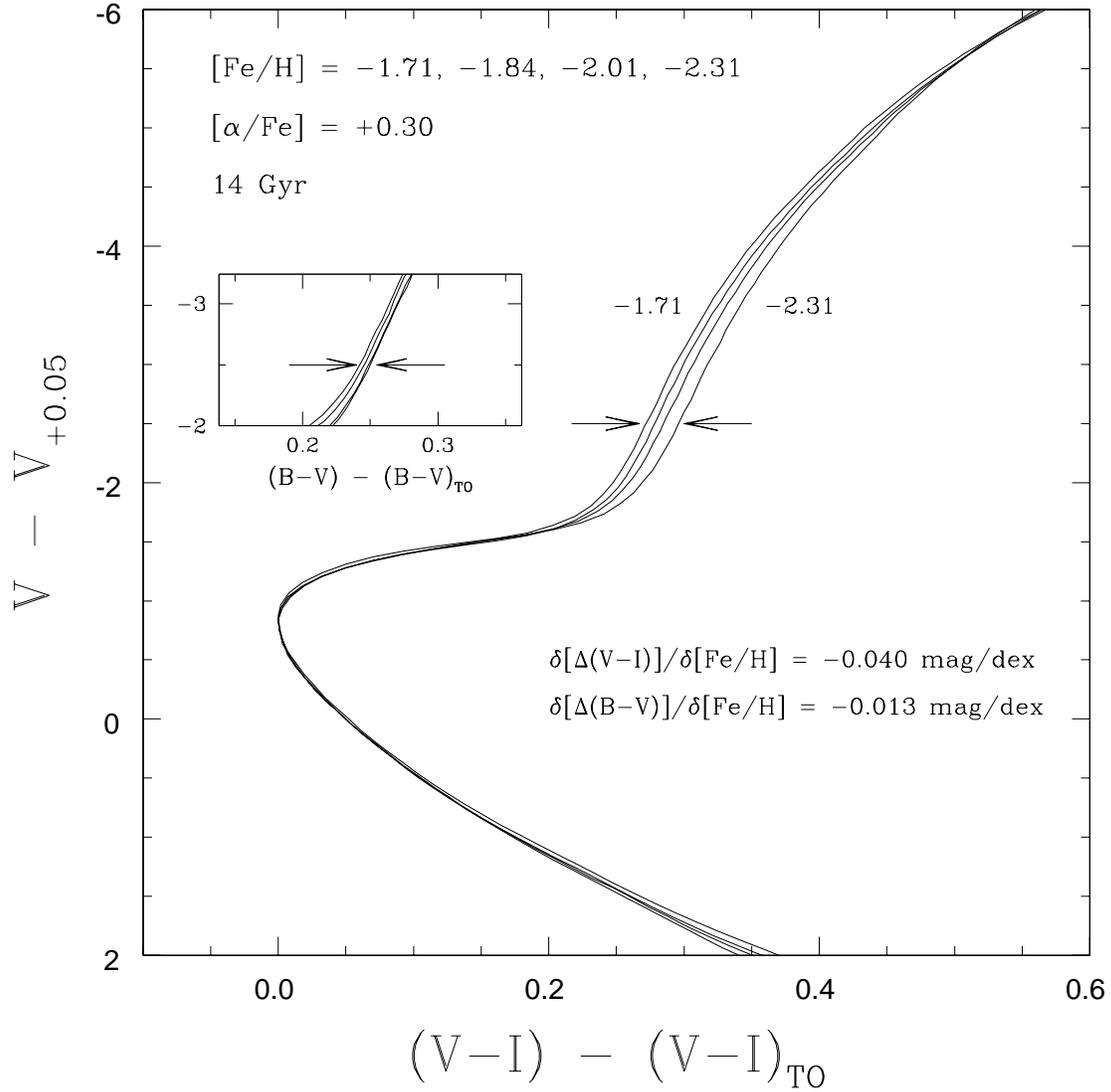}
\caption
{A plot of $(V-I) - (V-I)_{TO}$ versus $V - V_{+0.05}$
for model isochrones with [Fe/H] = $-$1.71, $-$1.84, $-$2.01, and $-$2.31
at the age of 14 Gyr (Bergbusch \& VandenBerg 2001), showing metallicity
dependence of the relative age estimates using the method recommended by
VandenBerg, Bolte, \& Stetson (1990).
The inset of the Figure shows a plot of $(B-V) - (B-V)_{TO}$ versus
$V - V_{+0.05}$ using the same model isochrones, where the metallicity effect
in the relative age estimates is smaller.}
\end{figure}

\begin{figure}
\epsscale{1}
\figurenum{7}
\plotone{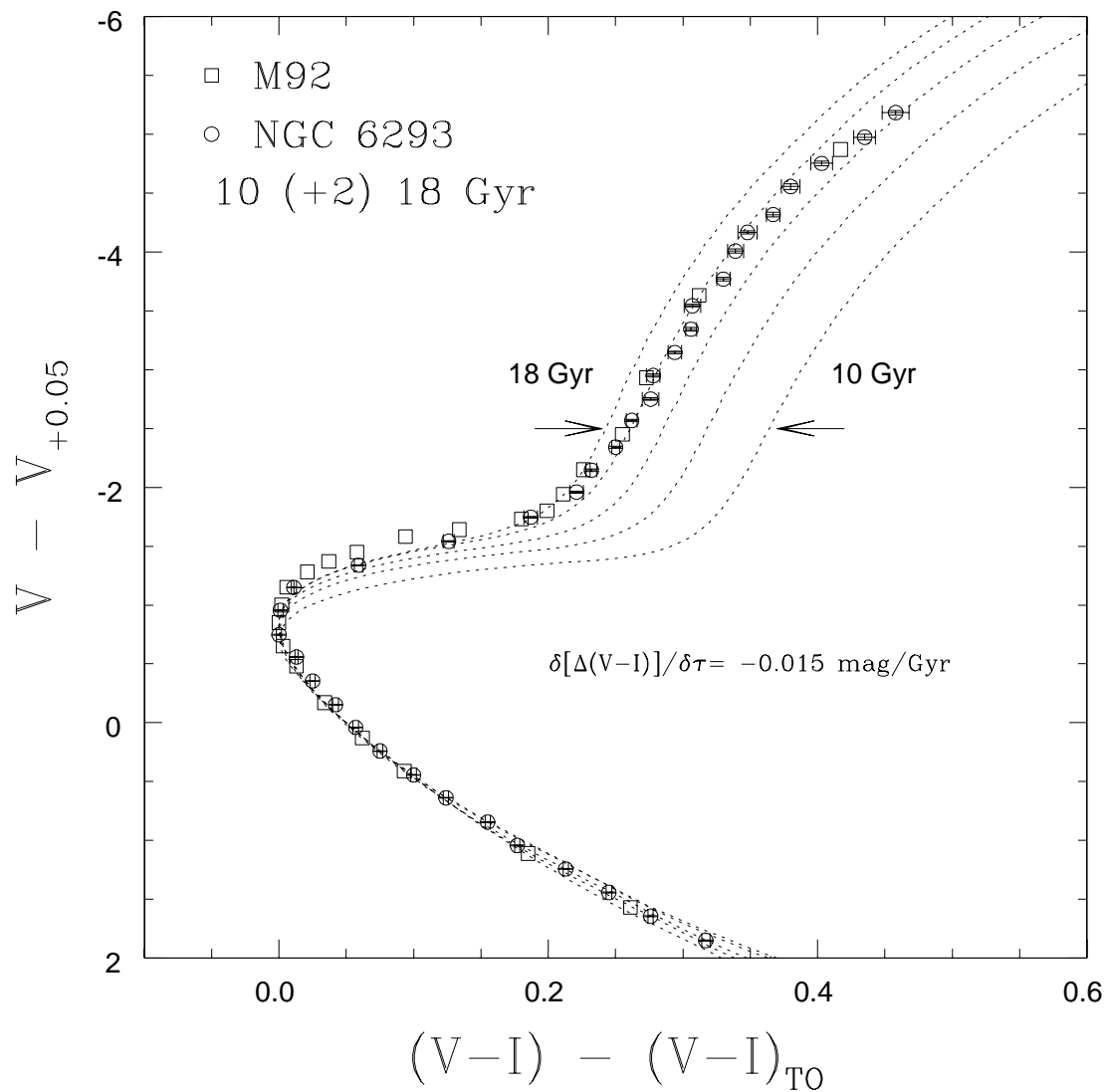}
\caption
{Plot of $(V-I) - (V-I)_{TO}$ versus $V - V_{+0.05}$ for fiducial sequences
of NGC~6293 and M92. The fiducial sequence of NGC~6293 is represented by
open circles and that of M92 by open squares.
The model isochrones by Bergbusch \& VandenBerg (2001)
were adopted in the plot.}
\end{figure}

\begin{figure}
\epsscale{1}
\figurenum{8}
\plotone{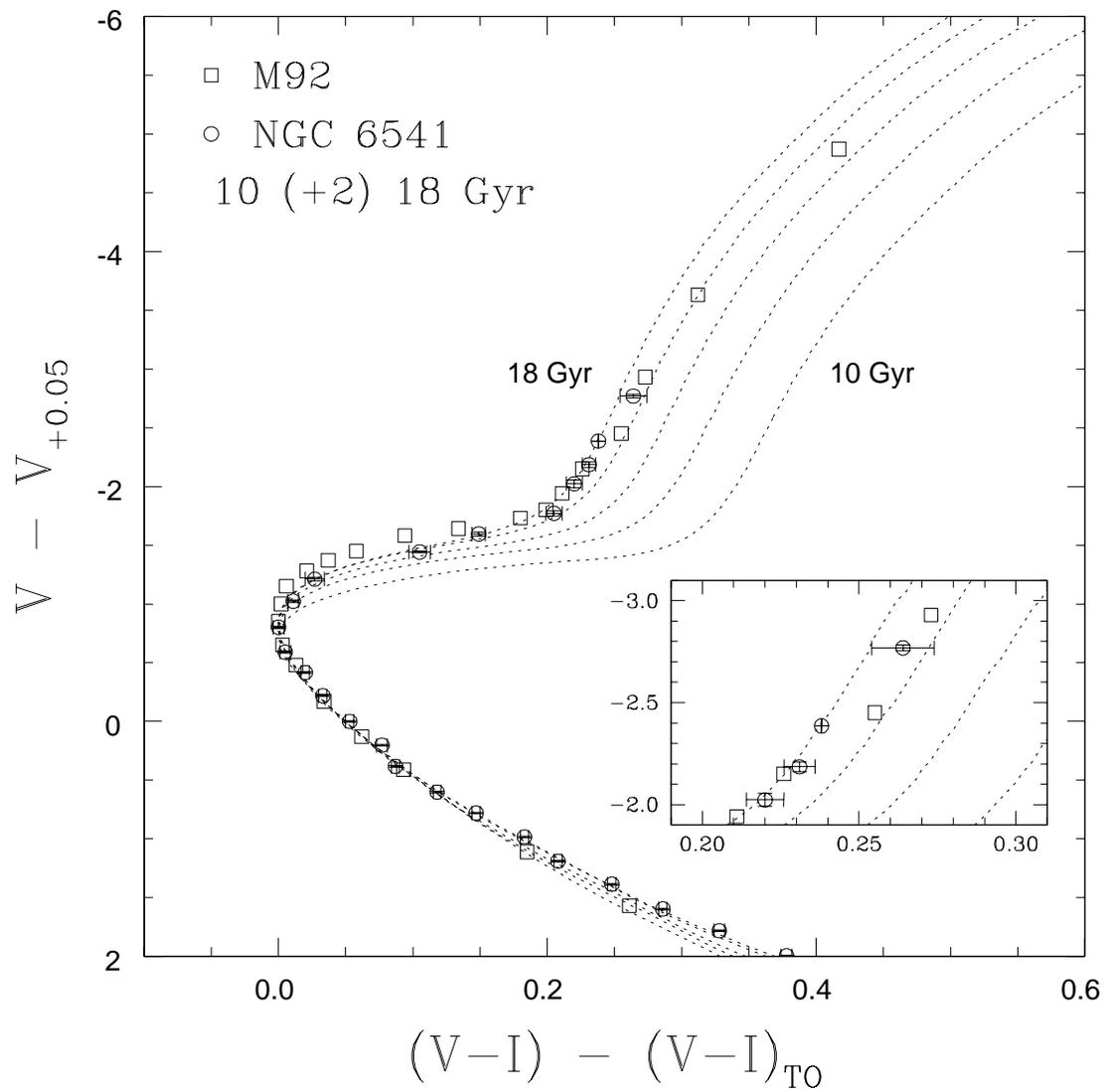}
\caption
{A plot of $(V-I) - (V-I)_{TO}$ versus $V - V_{+0.05}$
for fiducial sequences of NGC~6541 and M92.
The fiducial sequence of NGC~6541 is represented by open circles
and that of M92 by open squares.}
\end{figure}

\begin{figure}
\epsscale{1}
\figurenum{9}
\plotone{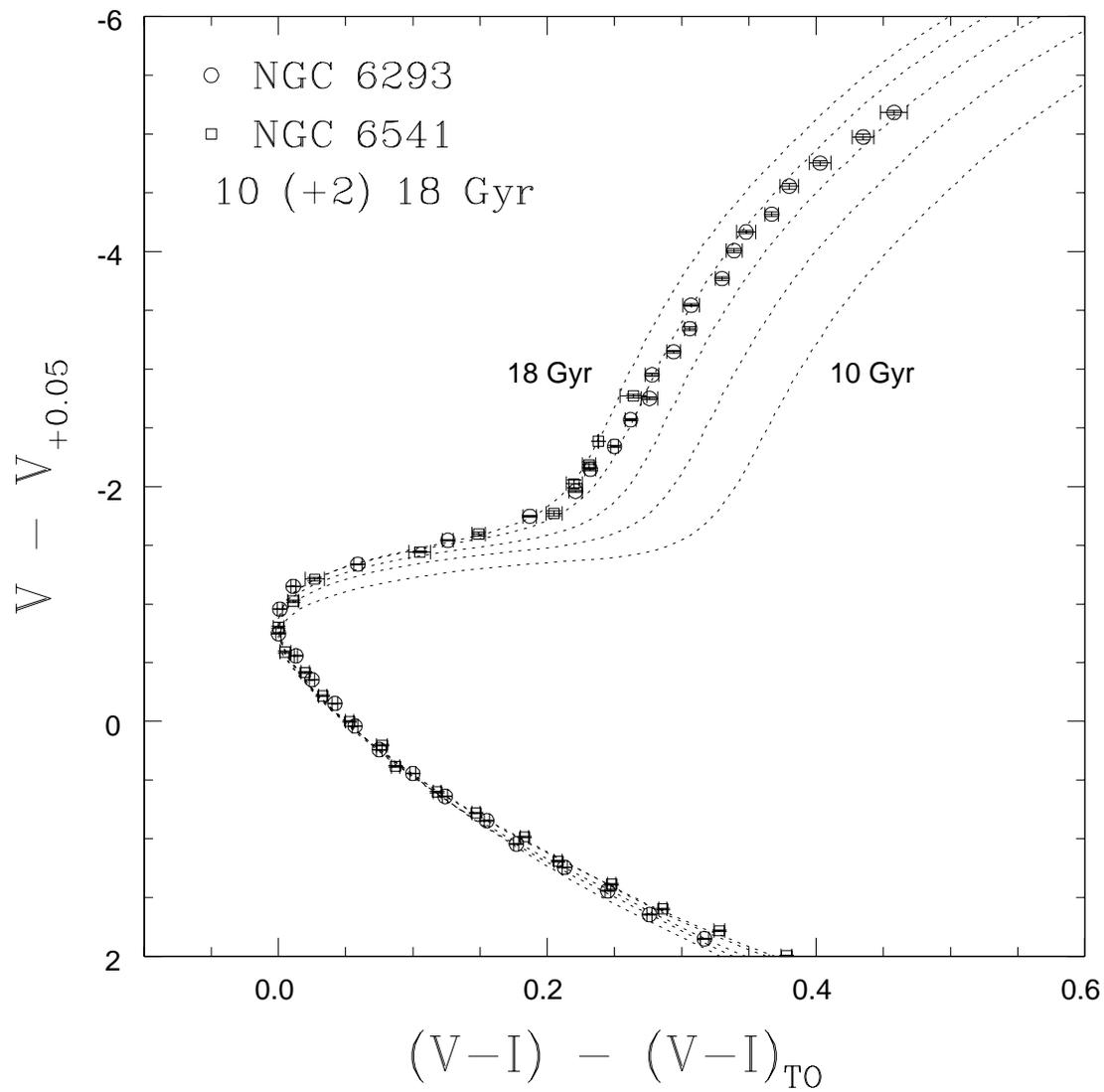}
\caption
{A plot of $(V-I) - (V-I)_{TO}$ versus $V - V_{+0.05}$
for fiducial sequences of NGC~6293 and NGC~6541.
The fiducial sequence of NGC~6293 is represented by filled circles
and that of NGC~6541 by open squares.}
\end{figure}

\begin{figure}
\epsscale{1}
\figurenum{10}
\plotone{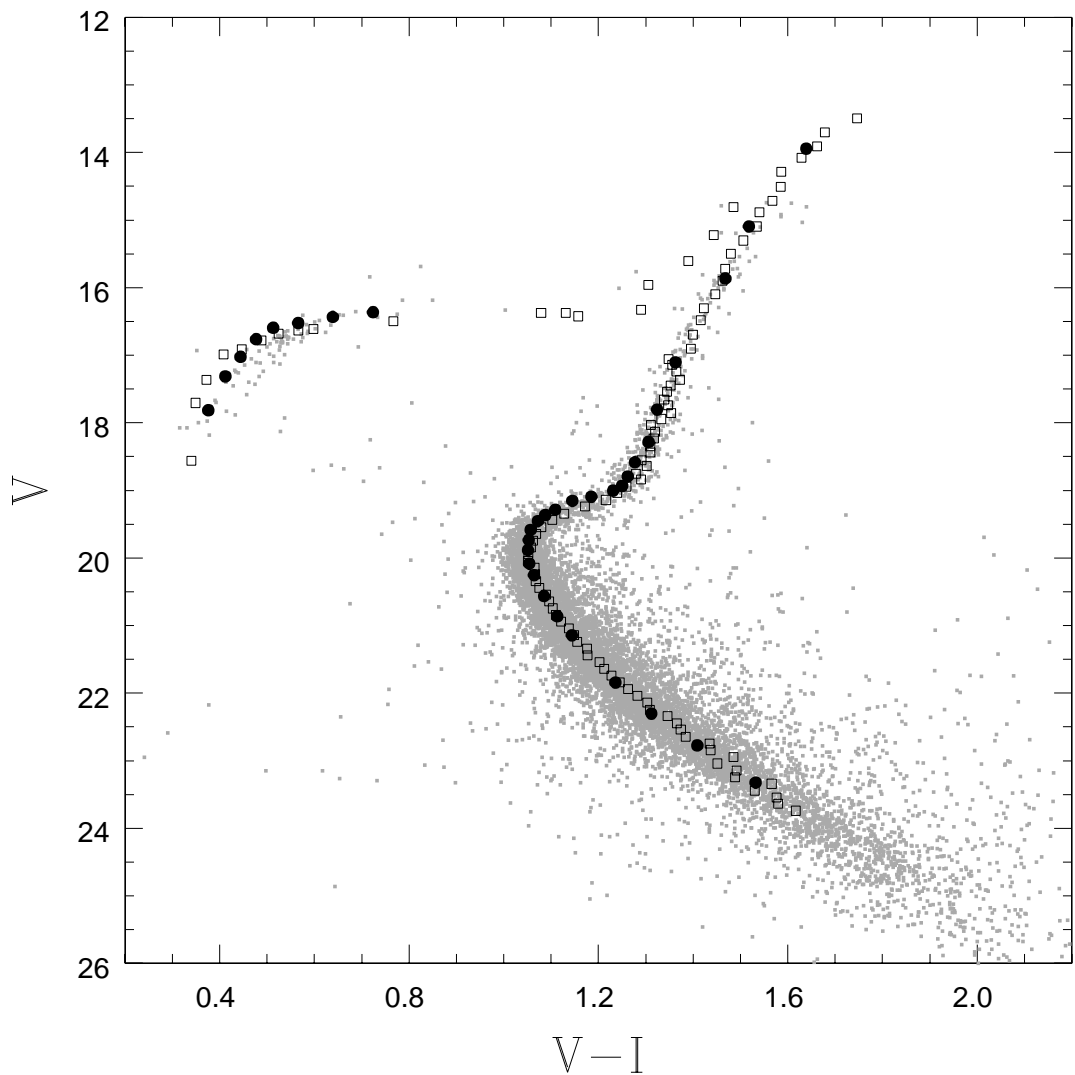}
\caption
{CMD of NGC~6293 along with the fiducial sequences of M92 and NGC~2419
by matching clusters' [$(V-I)_{TO}$, $V_{+0.05}$] points.
The filled circles are for M92 fiducial sequences of Johnson \& Bolte (1997)
and the open squares are for NGC~2419 fiducial sequence of
Harris et al.\ (1997).
In the Figure, we adopt $\delta V$ = 1.193 mag and $\delta (V-I)$ = 0.496 mag
for M92 and $\delta V$ = $-$4.008 mag and $\delta (V-I)$ = 0.374 mag
for NGC~2419, in the sense of NGC~6293 minus M92 or NGC~2419.}
\end{figure}

\begin{figure}
\epsscale{1}
\figurenum{11}
\plotone{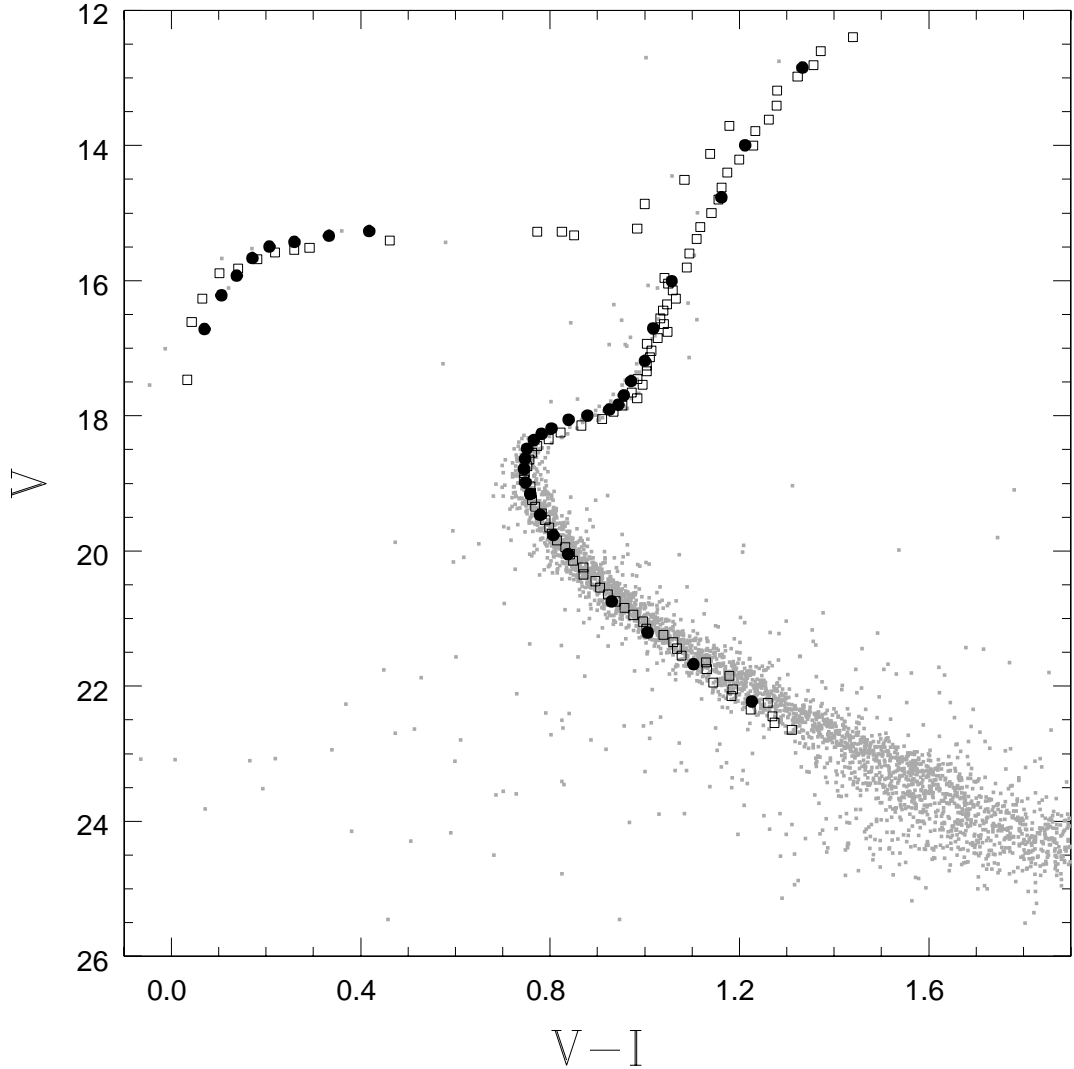}
\caption
{The CMD of NGC~6541 along with the fiducial sequences of M92 and NGC~2419
by matching clusters' [$(V-I)_{TO}$, $V_{+0.05}$] points.
The filled circles are for the M92 fiducial sequence of Johnson \& Bolte (1997)
and the open squares are for the NGC~2419 fiducial sequence of
Harris et al.\ (1997).
In the Figure, we adopt $\delta V$ = 0.096 mag and $\delta (V-I)$ = 0.190 mag
for M92 and $\delta V$ = $-$5.105 mag and $\delta (V-I)$ = 0.068 mag
for NGC~2419, in the sense of NGC~6541 minus M92 or NGC~2419.}
\end{figure}

\begin{figure}
\epsscale{1}
\figurenum{12}
\plotone{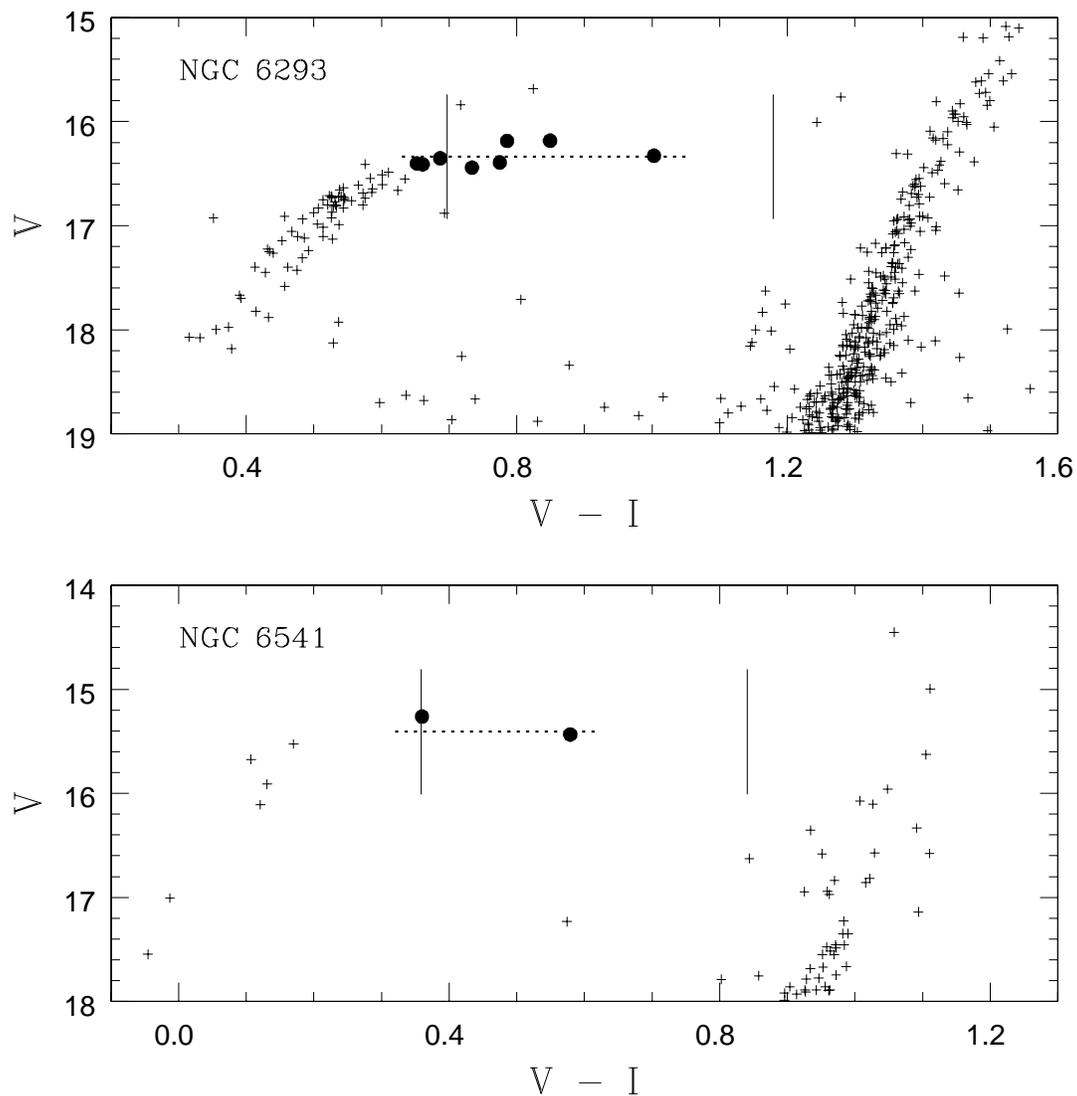}
\caption
{CMDs of the HB region in NGC~6293 and NGC~6541. The dotted lines
are the mean HB magnitude of the clusters and the filled circles
are the HB stars used in the $V_{HB}$ magnitude calculations.
We obtained $V_{HB}$ = 16.337 $\pm$ 0.036 mag (8 stars) for NGC~6293 and
$V_{HB}$ = 15.348 $\pm$ 0.087 mag (2 stars) for NGC~6541.
The vertical solid lines represent the first harmonic blue edge and
the fundamental red edge of M3 RR Lyrae variables (Carretta et al.\ 1998).}
\end{figure}

\begin{figure}
\epsscale{1}
\figurenum{13}
\plotone{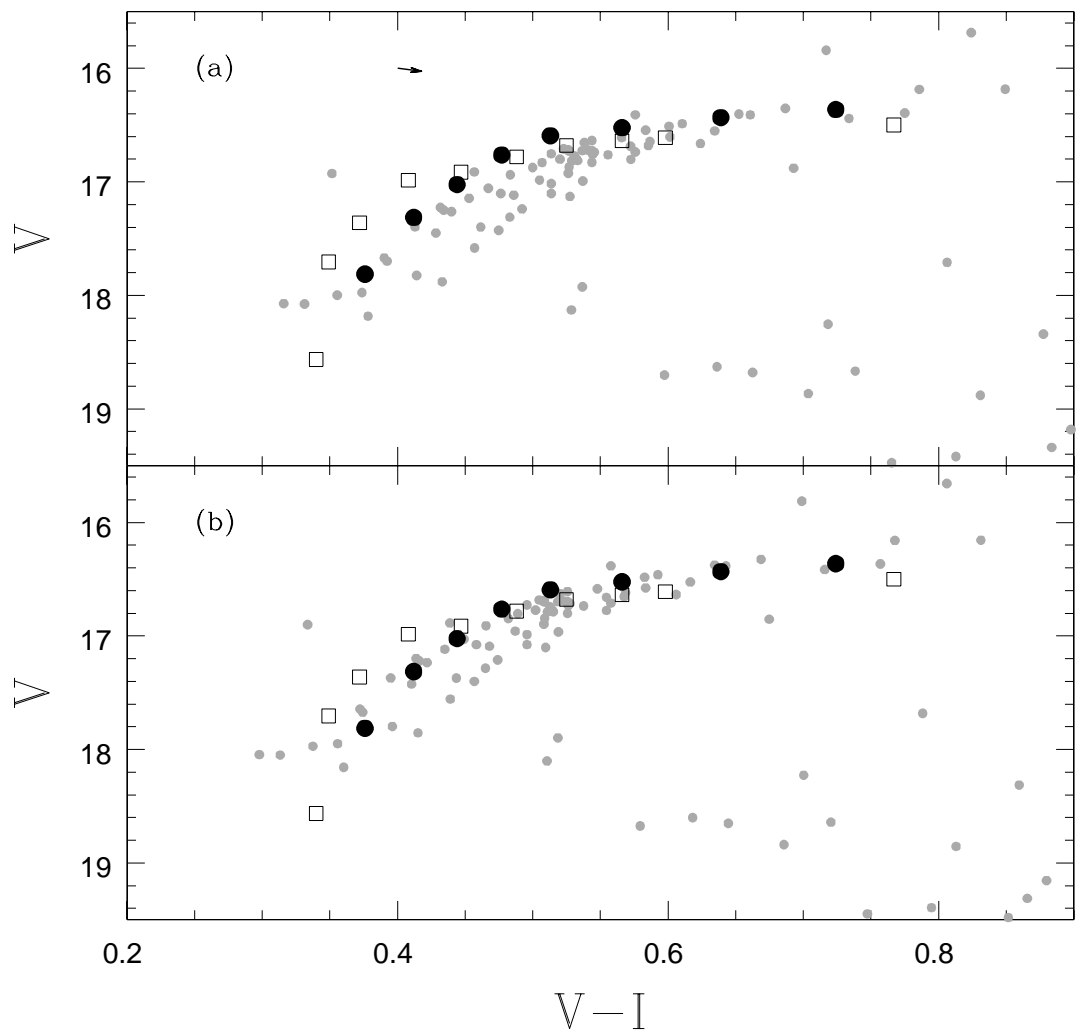}
\caption
{CMDs of the HB region in NGC~6293 without the differential reddening
correction with temperature (a) and after applying the differential reddening
correction (b). NGC~6293 HB stars are represented by gray dots and
HB fiducial sequences for M92 and NGC~2419 are represented by
filled circles and open squares, respectively.
The differential reddening correction vector due to the difference
in temperature of the source is indicated by an arrow in (a).
The location of HB stars is defined slightly better with
the differential reddening correction in (b).}
\end{figure}

\begin{figure}
\epsscale{1}
\figurenum{14}
\plotone{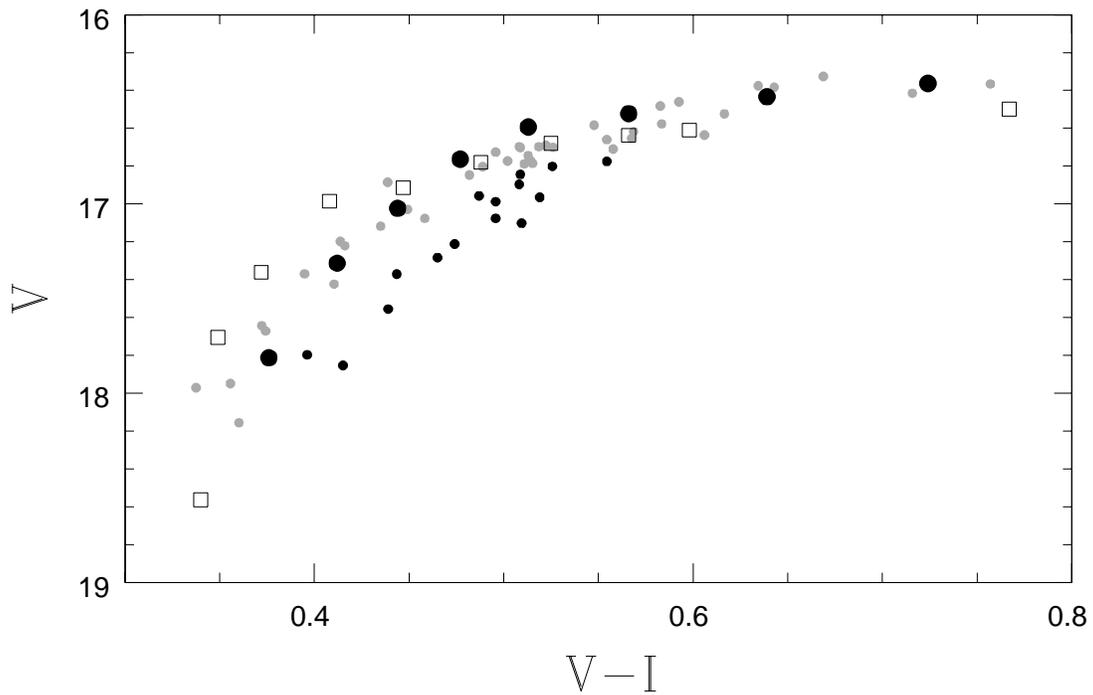}
\caption
{The CMD of the HB region in the field NGC~6293-F1. The M92 HB fiducial
sequence is represented by filled circles and that of NGC~2419 by
open squares. The HB stars with normal magnitude in NGC~6293 
are represented by gray dots and the {\em apparent} under-luminous HB stars
by black dots. In the plot, we adopt the temperature effect corrected
magnitudes and colors from Figure~13b.}
\end{figure}

\begin{figure}
\epsscale{1}
\figurenum{15}
\plotone{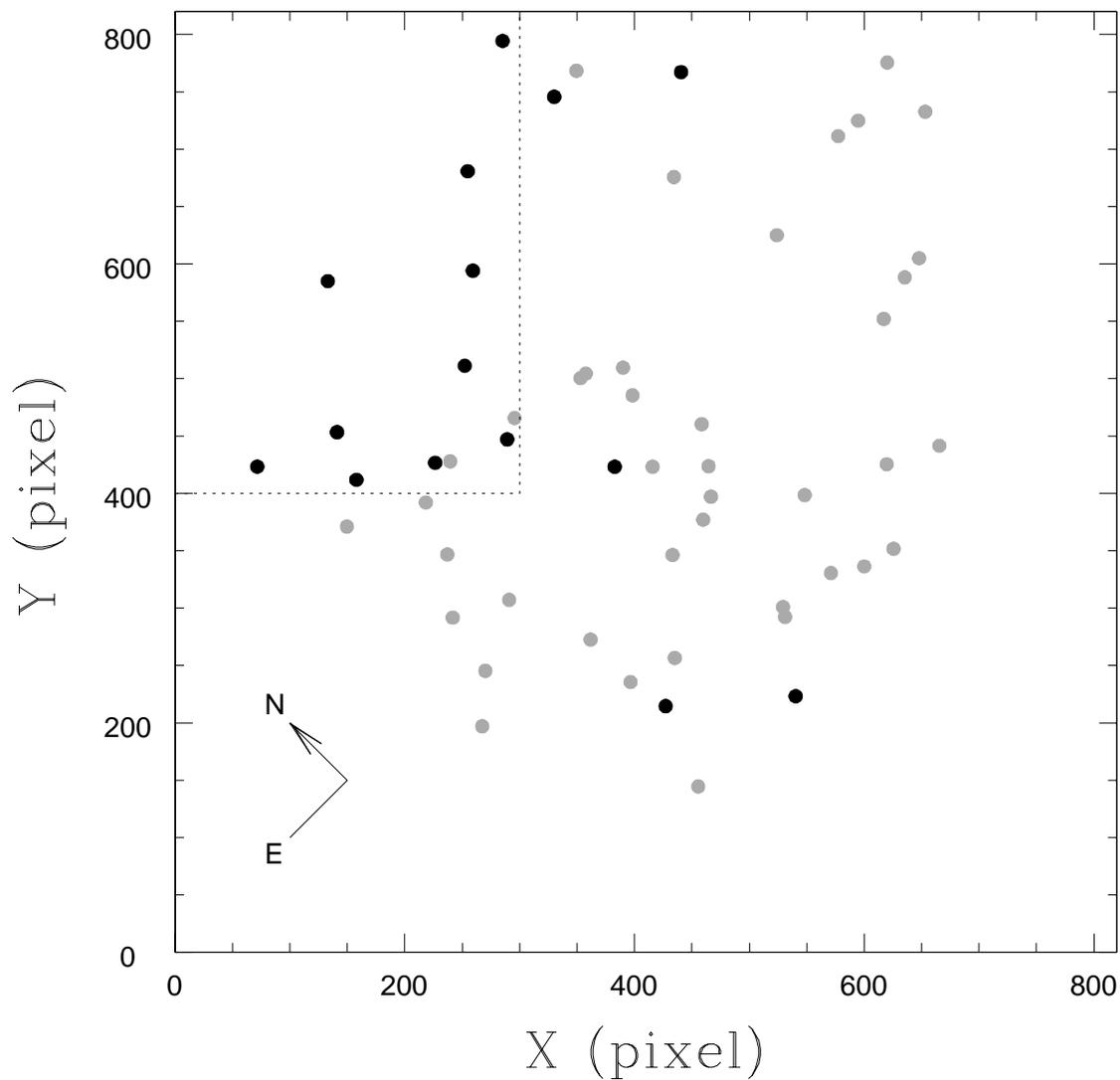}
\caption
{The spatial distribution of HB stars in the field NGC~6293-F1.
The HB stars with normal magnitude in NGC~6293 are represented by gray dots
and the {\em apparent} under-luminous HB stars by black dots.
Note that the most under-luminous HB stars are located in the upper left
corner (the north of the cluster).}
\end{figure}

\begin{figure}
\epsscale{1}
\figurenum{16}
\plotone{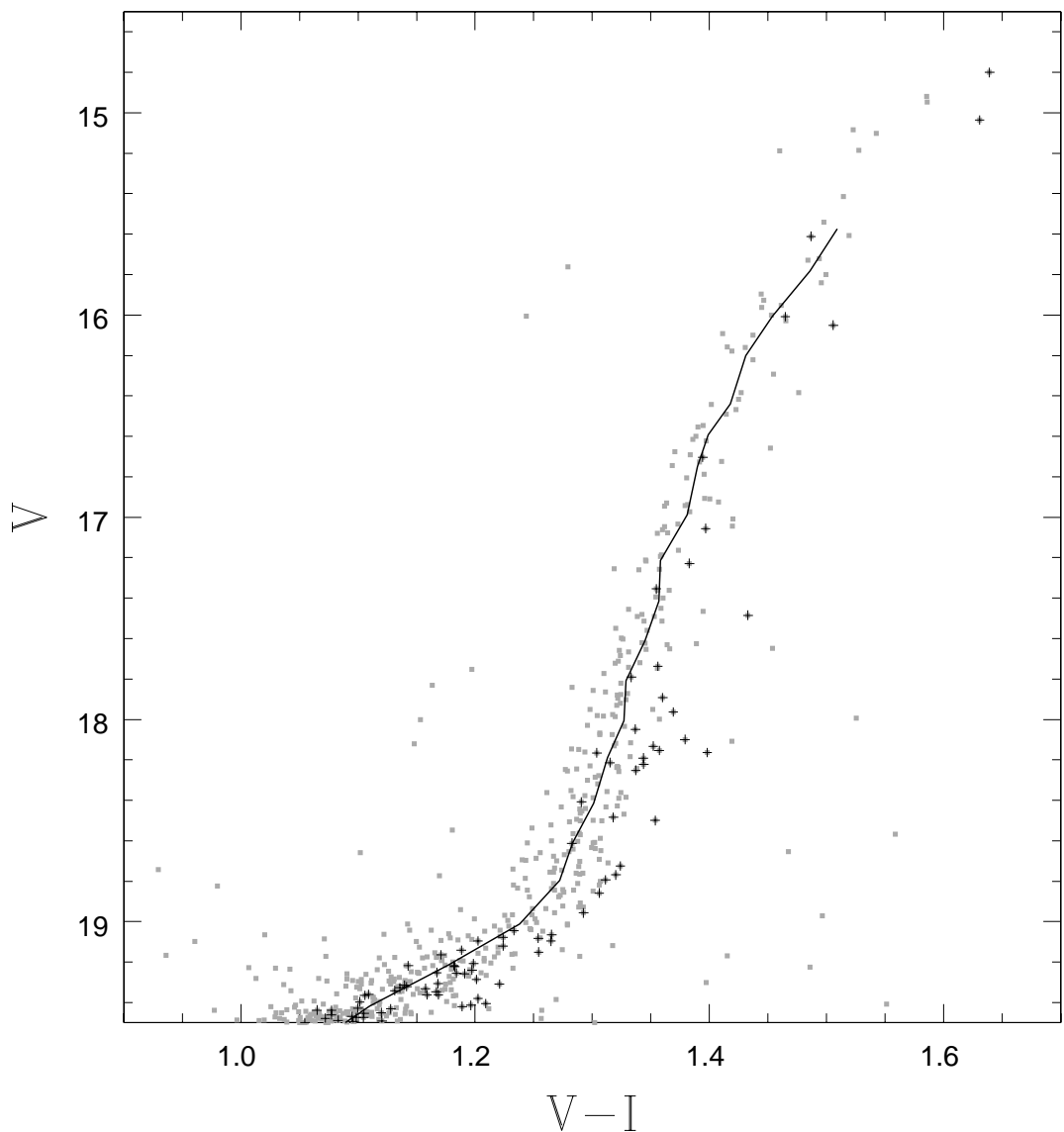}
\caption
{The CMD of RGB stars (plus signs) detected with $X$ $\leq$ 300
and $Y$ $\geq$ 400 in the field NGC~6293-F1
(inside the dotted lines in Figure~15), superposed on the CMD of stars
(gray dots) detected in the rest of the field.
The solid line represents the NGC~6293 RGB fiducial sequence in Table~\ref{tab4}.
The RGB stars detected in this region (where the under-luminous HB stars are
found) appear to be more reddened than the RGB stars in other region
of the cluster.}
\end{figure}

\begin{figure}
\epsscale{1}
\figurenum{17}
\plotone{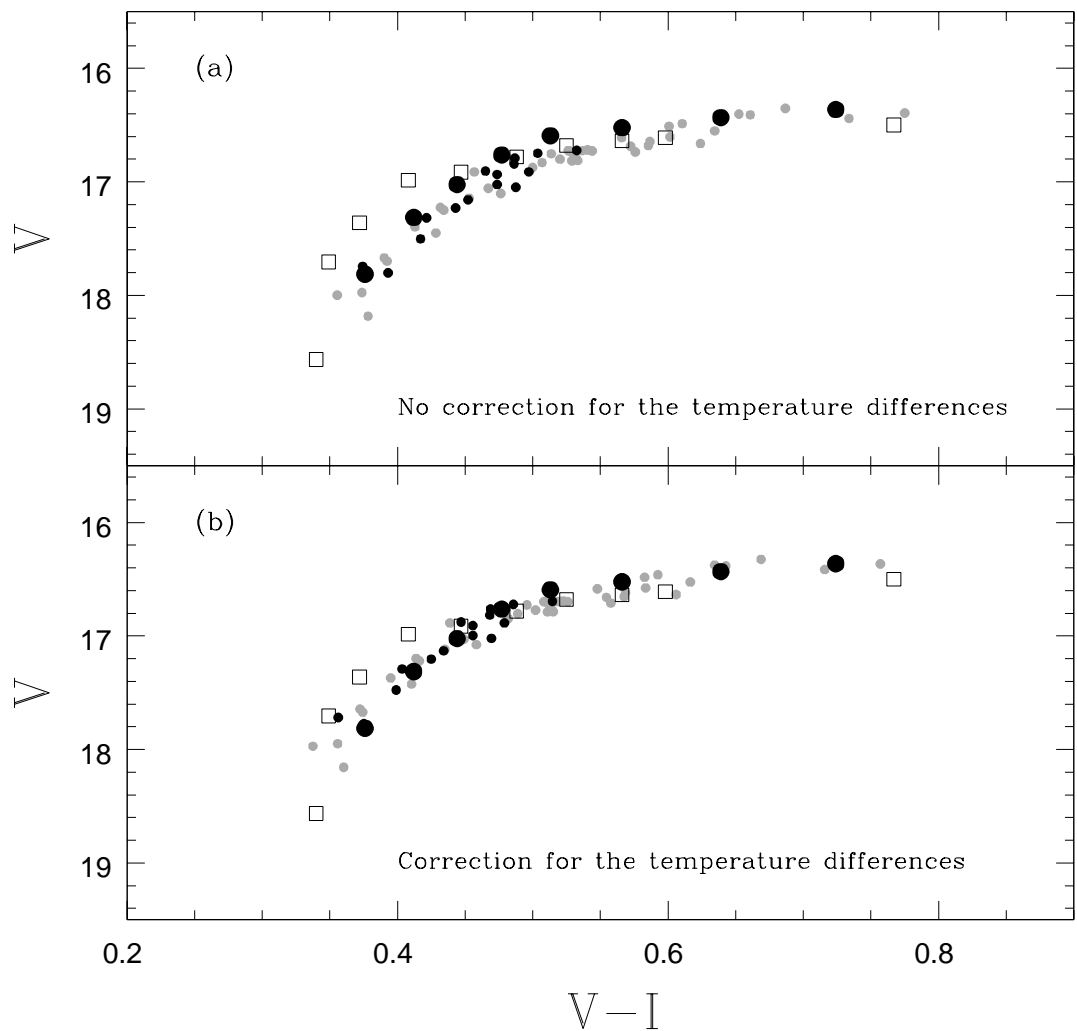}
\caption
{The CMD of HB stars in the field NGC~6293-F1 applying the differential
correction, $\Delta(V-I)$ = 0.04 mag and  $\Delta V$ = 0.08 mag, for
the under-luminous HB stars. Legends are the same as in Figure~14.
In (b), we also apply a reddening correction due to the difference
in the temperature of the source, $\Delta (V-I)$ = 0.02 mag and
$\Delta V$ = 0.03 mag.}
\end{figure}

\begin{figure}
\epsscale{1}
\figurenum{18}
\plotone{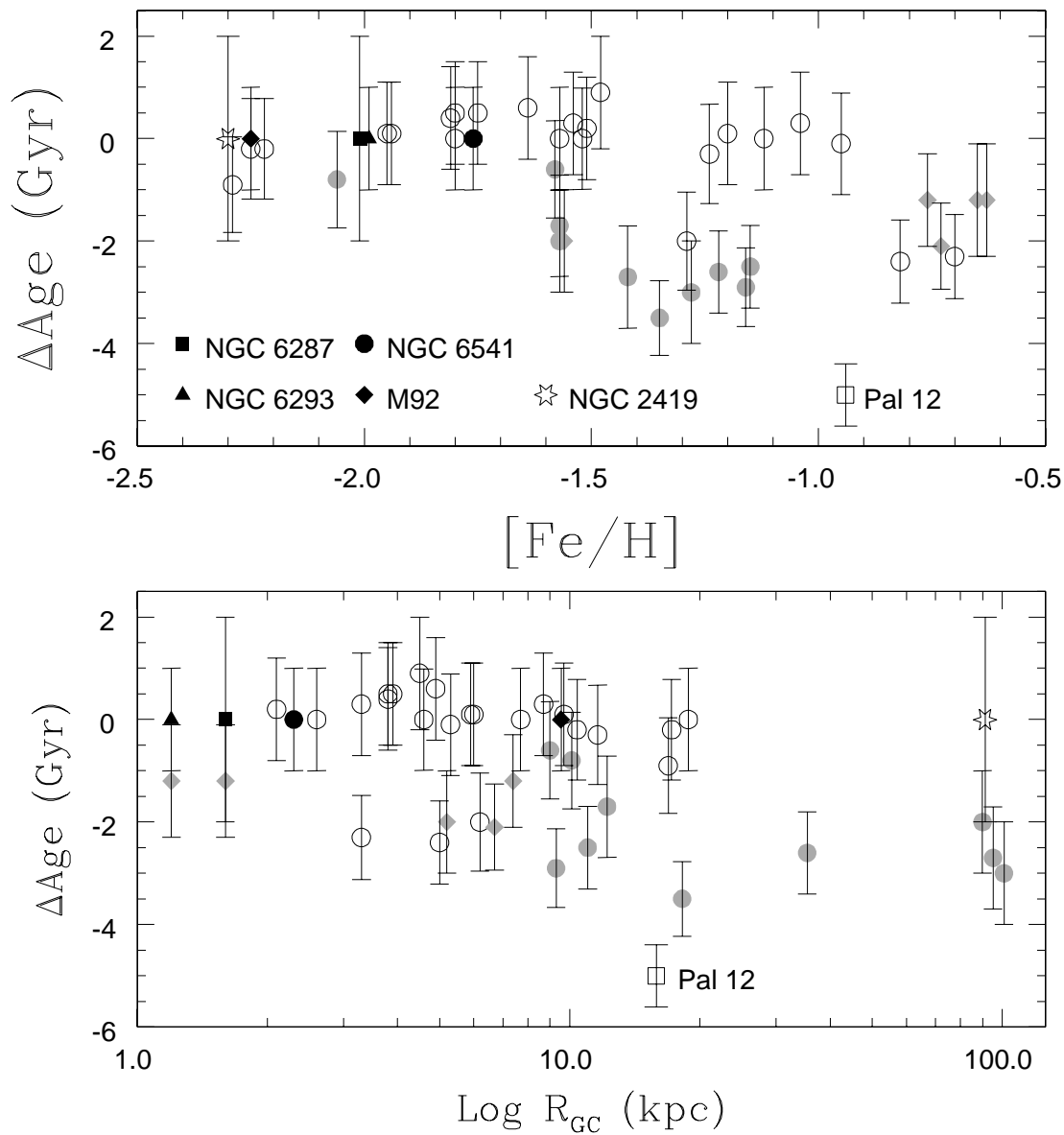}
\caption
{Comparisons of relative ages of the globular clusters as a
function of [Fe/H] and the Galactocentric distance $R_{GC}$.
The old halo globular clusters are presented by open circles,
the younger halo clusters by gray circles, and the thick disk clusters
by gray diamonds, and Pal~12, which is a member of the Sgr dSph,
by open squares. The ages of the globular clusters do not vary with
the Galactocentric distance or the metallicity with [Fe/H] $\leq$ $-$1.0.}
\end{figure}

\end{document}